\documentclass[10pt]{article}
\usepackage[utf8]{inputenc}
\usepackage{amsmath}
\usepackage{graphicx,tabularx}
\usepackage{amsfonts,amsthm,eucal,amssymb,amscd}
\usepackage{color}
\usepackage{epstopdf}
\usepackage{upgreek}
\usepackage{bbold} 
\usepackage{glossaries}
\usepackage{amssymb}
\usepackage{tikz}
\usepackage[all]{xy}
\usetikzlibrary{matrix}
\usepackage{graphicx}

\newtheorem{definition}{Definition}

\begin{document}

\title{Generalized Probabilities in Statistical Theories}
\author{{\sc Federico Holik}$^{1}$$^{\ast}$, {\sc C\'esar Massri}$^{2}$, {\sc Angelo Plastino}$^{1}$ {\sc Manuel S\'{a}enz}$^{2}$}

\maketitle

\begin{center}

\begin{small}
$^{1}$ Instituto de F\'{\i}sica (IFLP-CCT-CONICET), Universidad
Nacional de La Plata, C.C. 727, 1900 La Plata, Argentina\\
$^{2}$ Departamento de Matem\'{a}tica, Universidad CAECE (1084), Buenos Aires, Argentina;\\
$^{3}$ International Center for Theoretical Physics, UNESCO, (34151)
Trieste, Italy;\\
$^{\ast}$ holik@fisica.unlp.edu.ar
\end{small}
\end{center}

\vspace{1cm}

\begin{abstract}
We discuss different formal frameworks for the description of
generalized probabilities in statistical theories. We analyze the
particular cases of probabilities appearing in classical and quantum
mechanics and the approach to generalized probabilities based on
convex sets. We argue for considering quantum probabilities as the
natural probabilistic assignments for rational agents dealing with
contextual probabilistic models. In this way, the formal structure
of quantum probabilities as a non-Boolean probabilistic calculus is
endowed with a natural interpretation.
\end{abstract}

\begin{small}
\centerline{\em Keywords: quantum probability; lattice theory;
information theory; classical probability; Cox's approach}
\end{small}

\section{Introduction}
In the year 1900, the great mathematician David Hilbert presented a
famous list of problems at a Conference in Paris. Hilbert suggested
that the efforts of the mathematicians in the years to come should
be oriented in the solution of these problems. The~complete list was
published later~\cite{HilbertProblems-1902}. Remarkably, one of
these problems was dedicated to the axiomatic treatment of
probability theory and physical theories. In~Hilbert's own words
(\cite{HilbertProblems-1902}, p.~454):

\begin{quote}
``The investigations on the foundations of geometry suggest the
problem: To treat in the same manner, by~means of axioms, those
physical sciences in which mathematics plays an important part; in
the first rank are the theory of probabilities and mechanics. As~to
the axioms of the theory of probabilities, it seems to me desirable
that their logical investigation should be accompanied by a rigorous
and satisfactory development of the method of mean values in
mathematical physics, and~in particular in the kinetic theory of
gases.''
\end{quote}

After a series of preliminary investigations by many researchers
(see, for example~\cite{Rocchi}), an~axiomatization of probability
theory was finally presented in the 1930s by Andrey
Kolmogorov~\cite{KolmogorovProbability}. This contribution, which
can be considered as the foundation of modern probability theory, is
based on measure theory. Indeed, in~Kolmogorov's axiomatic
treatment, probability is considered as a measure defined over a
suitable collection of events, organized as a \emph{sigma-algebra}
(which
is found to be also a \emph{Boolean lattice}). 
His list of axioms allows the description of many examples of
interest and was considered as a reasonable fulfillment of Hilbert's
program for probability~theory.

Hilbert himself dedicated great efforts to solve his sixth problem.
His contributions were influential in the development of Relativity
Theory, and~he also contributed to the development of Quantum
Mechanics. Indeed, Quantum Mechanics acquired its rigorous axiomatic
formulation after a series of papers by Hilbert, J. von Neumann, L.
Nordheim, and~E. P. Wigner~\cite{vN-Hilbert-Nordheim}. It can be
said that its definitive form was accomplished in the book of von
Neumann~\cite{vN}. This axiomatic approach was extended to the
relativistic setting in subsequent years (see, for
example,~\cite{HaggLQP,Wightman-1964}; see~\cite{HalvorsonARQFT} for
a more updated exposition of the algebraic approach; and for a
rigorous formulation of quantum statistical mechanics,
see~\cite{Bratteli}).

However, the advent of Quantum Mechanics presented a model of
probabilities that had many peculiar features. R. P. Feynman stated
this clearly in~\cite{FeynmannProbability}, p.~533:

\begin{quote}
``I should say, that in spite of the implication of the title of
this talk the concept of probability is not altered in quantum
mechanics. When I say the probability of a certain outcome of an
experiment is \emph{p}, I mean the conventional thing, that is,
if~the experiment is repeated many times one expects that the
fraction of those which give the outcome in question is roughly $p$.
I will not be at all concerned with analyzing or defining this
concept in more detail, for~no departure of the concept used in
classical statistics is required. What is changed, and~changed
radically, is the method of calculating probabilities.''
\end{quote}

What is the meaning of Feynman's words? Feynman tells us that the
way of computing frequencies is not altered in quantum mechanics:
the real numbers yielded by Born's rule can be tested in the lab in
the usual way. However, the method for \emph{computing}
probabilities has changed in a radical way. As~put
in~\cite{Fuch-Schack-2013}, this can be rephrased as follows: the
radical change has to do with the recipe that quantum mechanics
gives us for calculating new probabilities from old. The~radical
change mentioned by Feynman lies behind all the astonishing features
of quantum phenomena. This was recognized very quickly as a
non-classical feature. These peculiarities and the formal aspects of
the probabilities involved in quantum theory have been extensively
studied in the
literature~\cite{Redei-Summers2006,Davies-Lewies,Srinivas,mackey57,Gudder-StatisticalMethods,mackey-book,Ehti-2015,HolikRingsOfOperators}.
We  refer to the probabilities related to quantum phenomena as
\emph{quantum probabilities} (QP). Accordingly, we  refer to
probabilities obeying Kolmogorov's axioms as \emph{classical
probabilities} (CP).

In this paper, we  discuss the formal structure of quantum
probabilities as measures over a non-Boolean algebra. We focus on a
crucial aspect of quantum probabilities---namely, that there exists
a major structural difference between classical states and
quantum~states:

\begin{itemize}
\item States of classical probabilistic systems can be suitably
described by Kolmogorovian measures. This is due to the fact that
each classical state defines a measure in the Boolean sigma-algebra
of measurable subsets of phase space.
\item Contrarily to classical states, quantum states cannot be reduced
to a single Kolmogorovian measure. A~density operator representing a
quantum state defines a measure over an orthomodular lattice of
projection operators, which contains (infinitely many) incompatible
maximal Boolean subalgebras. These represent different and
complementary---in the Bohrian sense---experimental setups. The~best
we can do is to consider a quantum state as a family of
Kolmogorovian measures, pasted in a harmonic
way~\cite{HolikQIC-2016}; however,~there is no joint (classical)
probability distribution encompassing all possible contexts.
\end{itemize}

We discuss the above mentioned differences in relation to quantum
theory as a non-Kolmogorovian probability calculus. This calculus
can be considered as an extension of classical measure theory to a
non-commutative setting (see, for
example~\cite{Redei-Summers2006,mikloredeilibro}; see
also~\cite{Hamhalter-QuantumMeasureTheory} for a study of quantum
measure theory). In~this way, the~axiomatization of probabilities
arising in QM (and more general probabilistic models) can be viewed
as a continuation of the Hilbert's program with regard to
probability theory. We argue that the probabilities in generalized
probabilistic models can be interpreted, in~a natural way, in~terms
of reasonable expectations of a rational agent facing event
structures that may define different and incompatible contexts. This
allows us to understand other related notions, such as random
variables and information measures, as~natural generalizations of
the usual ones.

Kolmogorov's approach to probability theory is not the only one.
In~the attempts to establish foundations for probability, we have to
mention the works of de Finetti~\cite{deFinetti} and R. T.
Cox~\cite{CoxLibro,CoxPaper} (in connection with R. T. Cox works,
see also~\cite{Jaynes2}). For~a detailed and accessible study of the
history of probability theory and its interpretations, we refer the
reader to the Apendix of~\cite{Rocchi}. In~this paper, we pay
special attention to Cox's approach and make use of its extension to
the quantum realm~\cite{Holik-Plastino-Saenz-2012}. Cox's approach
is based on a study of the measure functions compatible with the
algebraic properties of the logic of a rational agent trying to make
probable inferences out of the available data. Different variants of
this approach have been used to describe probabilities in
QM~\cite{Caticha-99,GoyalKnuthSkilling,Knuth-2004a,Knuth-2004b,Knuth-2005a,Knuth-2005b,Symmetry,Holik-Plastino-Saenz-2012}.

In~\cite{Holik-Plastino-Saenz-2012}, it is shown that the peculiar
features of QP arise whenever the lattice of propositions of Cox's
approach is replaced by a non-distributive one. As~is well known,
the~standard quantum-logical approach to QM characterizes this
theory using a lattice-theoretical framework in which the lattices
are
orthomodular~\cite{BvN,dallachiaragiuntinilibro,mikloredeilibro,belcas81,vadar68,vadar70,svozillibro,HandbookofQL,jauch,
piron,kalm83,kalm86,greechie81,giunt91,pp91,dvupulmlibro,aertsdaub1,aertsjmp83,aertsrepmathphys84a,aertsjmp84b}.
In~\cite{Holik-Plastino-Saenz-2012}, and it is shown that, when
Cox's method is applied to the orthomodular lattice of projections
of a Hilbert space, QP are~derived.

We remark that generalized probabilities can also be studied in what
has been called the Convex Operational Models (COM) approach~\cite{Beltrametti.Varadarajan-2000,
Barnum-Wilce-2009,Barnum-Wilce-2010,Barnum-PRL,Barnum-Toner,barnum,Entropy-generalized-II,Barret-2007,Perinotti-2011,GenealizedTeleportationBarnum,
Holik-Plastino-Massri}. These are called ``generalized probabilistic
theories'' (GPTs) or~simply, ``operational models'' (see also
\cite{mackey-book}).

Different mathematical frameworks are used to describe GPTs. Here,
we focus on the most important ones, paying attention to how they
can be inter-translated (when possible). In~the COM approach,
the~properties of the systems studied and their associated
probabilities are encoded in a geometrical way in the form of a
convex set and its space of observable magnitudes. The~quantum
formalism and many aspects of quantum information theory (such as
entanglement, discord, and many information protocols) can be
suitably described using this
approach~\cite{barnum,Entropy-generalized-II,Barret-2007,Barnum-Toner,Holik-Plastino-Massri,GenealizedTeleportationBarnum,Barnum-PRL,Barnum-Wilce-2009,Barnum-Wilce-2010,Perinotti-2011}.
Non-linear generalizations of QM were studied using the convex
approach in~\cite{Mielnik-68,Mielnik-69,Mielnik-74}.

{It is important to understand the relations between the different
formulations of GPTs. For~example, the~measures over complete
orthomodular lattices discussed in Section~\ref{s:Orthomodular} of
this work define GPTs (while an arbitrary GPT might be not be
describable in terms of a measure over an orthomodular lattice).
The~reason why models defined over lattices are so important is that
all relevant physical theories can be described in such a setting.
Indeed, all relevant physical models can be ultimately described
using von Neumann algebras, that are generated by their lattices of
projection operators (see, for
example~\mbox{\cite{Bratteli,HalvorsonARQFT,mikloredeilibro}}). This
is the case for classical statistical theories, standard quantum
mechanics, relativistic quantum mechanics, and quantum statistical
mechanics. As~an example (as we will discuss in
\mbox{Section~\ref{s:QuantumProbabilities}}), states of models of
quantum mechanics can be described as measures over the orthomodular
lattice of projection operators acting on a separable Hilbert space.
It is interesting, for~several reasons, to~study more general models
(that could describe, for~example, alternative physical theories).
However, there is always a trade off between generality and
particularities: if our models are too general, we can loose
valuable information about the geometric and algebraic structures
involved in relevant physical theories. On~the contrary, if~they are
too specific, we might loose information about the general road map
for exploration. It is our aim here to shed some light onto this
vast field of research, putting the focus on the idea that
Kolmogorov's framework can  certainly be generalized in a reasonable
and useful way.}

It is also important to mention that quantum-like probabilities have
been considered outside the quantum domain. Many probabilistic
systems behave in a contextual way, and~then, it is reasonable to
attempt to use quantum-like probabilities to described them, since
these are specially suited to deal with contextual behavior. This is
an exciting field of research that has grown intensively during
recent years (see, for
example~\cite{Khrennikov-Ubiquitous,Narens-Book,Narens-Paper-2014}).

We start by reviewing different approaches to CP, namely,
Kolmogorov's and Cox's, in~Section~\ref{s:CoxReview}. Next, we
discuss the formalism of QM in Section~\ref{s:TheFormalismOfQM},
emphasizing how it can be considered as a non-Boolean version of
Kolmogorov's theory. In~Sections~\ref{s:Orthomodular} and
\ref{s:COMpreliminaries}, we discuss generalizations using
orthomodular lattices and COMs, respectively. After~discussing
alternative approaches in Section~\ref{s:KnuthGoyal}, we present the
generalization of the Cox method to general non-distributive
lattices in Section~\ref{s:QMDerivation}. Finally, our  conclusions
are drawn in Section~\ref{s:Conclusions}. Given that lattice theory
is so central to the discussions presented here, we have included a
short review about its elementary notions in
Appendix~\ref{s:LatticeTheory}.

\section{Classical~Probabilities}\label{s:CoxReview}

This Section is devoted to \emph{classical probability theory} (CP).
However, what do we mean with this notion? There exists a vast
literature and tendencies disputing the meaning of CP. We will not
give a detailed survey of the discussion here; however,~we will
discuss two of the most important approaches to CP. These are the
one given by A. N. Kolmogorov~\cite{KolmogorovProbability} and the
one given by R. T. Cox~\cite{CoxLibro,CoxPaper}.

\subsection{Kolmogorov}\label{s:Kolmogorov}

Kolmogorov presented his axiomatization of classical probability
theory~\cite{KolmogorovProbability} in the 1930s. It can be
formulated as follows. Given an outcome set $\Omega$, let us
consider a $\sigma$-algebra $\Sigma$ of subsets of $\Omega$.
A~probability measure will be a function $\mu$ such~that

\begin{subequations}\label{e:kolmogorovian}
\begin{equation}\label{e:1a}
\mu:\Sigma\rightarrow[0,1]
\end{equation}

\noindent satisfying
\begin{equation}\label{e:1b}
\mu(\Omega)=1, \, \\
\end{equation}

\noindent {and, if~$I$ is a denumerable set of indices, for~any
pairwise disjoint family $\{A_{i}\}_{i\in I}$},
\begin{equation}\label{e:1c}
\mu(\bigcup_{i\in I}A_{i})=\sum_{i}\mu(A_{i})
\end{equation}

\end{subequations}

Conditions \eqref{e:1a}--\eqref{e:1c} are known as \emph{axioms of
Kolmogorov}~\cite{KolmogorovProbability}. The~triad
$(\Omega,\Sigma,\mu)$ is called a \emph{probability space}.
Probability spaces obeying Equations \eqref{e:1a}--\eqref{e:1c} are
usually referred as Kolmogorovian, classical, commutative, or
Boolean probabilities~\cite{Gudder-StatisticalMethods}, due to the
Boolean character of the $\sigma$-algebra in which they are~defined.

It is possible to show that, if $(\Omega,\Sigma,\mu)$ is a
Kolmogorovian probability space, the~\emph{inclusion--exclusion
principle} holds
\begin{equation}\label{e:SumRule}
\mu(A\cup B)=\mu(A)+\mu(B)-\mu(A\cap B)
\end{equation}

\noindent or, as~expressed in logical terms, by replacing ``$\vee$''
instead of ``$\cup$'' and ``$\cap$'' instead of ``$\wedge$'':
\begin{equation}\label{e:SumRuleLogical}
\mu(A \vee B)=\mu(A)+\mu(B)-\mu(A \wedge B)
\end{equation}

As remarked in~\cite{RedeiHandbook}, Equation~\eqref{e:SumRule} was
considered as crucial by von Neumann for the interpretation of
$\mu(A)$ and $\mu(B)$ as relative frequencies. If~$N_{(A\cup B)}$,
$N_{(A)}$, $N_{(B)}$, $N_{(A\cap B)}$ are the number of times for
each event to occur in a series of $N$ repetitions, then
\eqref{e:SumRule} trivially holds. Notice that
\eqref{e:SumRuleLogical} implies that:
\begin{equation}\label{e:SumRuleInequality}
\mu(A)+\mu(B)\leq\mu(A \vee B)
\end{equation}

The inequality \eqref{e:SumRuleInequality}  no longer holds in QM,
a~fact linked to its non-Boolean character (see, for
example~\cite{Gudder-StatisticalMethods}),
Section~\ref{s:RandomVariables}. Indeed, for~a suitably chosen state
and~events $A$ and $B$ (i.e., for~a non-commutative pair),
\eqref{e:SumRuleInequality} can be violated. If~$N(A\vee B)$,
$N(A)$, $N(B)$ and $N(A\wedge B)$ are the number of times for each
event to occur in a series of N repetitions, then the sum rule
should trivially hold (but it does not). This poses problems to a
relative-frequencies' interpretation of quantum probabilities (see,
for example, the discussion posed in~\cite{RedeiHandbook}). The~QM
example shows that non-distributive propositional structures give
rise to probability theories that appear to be very {\it different
from} those of Kolmogorov. Notwithstanding, it is important to
mention that some authors have managed to develop a relative
frequencies interpretations (see, for
example~\cite{Ballentine-1970}).

If all possible measures satisfying \eqref{e:1a}--\eqref{e:1c} are
considered as forming a set $\Delta(\Sigma)$ (with $\Sigma$ fixed),
then it is straightforward to show that it is convex. As~we shall
see below, it is a simplex, and~its form will be related to the
Boolean character of the lattice of classical~events.

\subsection{Random Variables and Classical~States}\label{s:RandomVariables}

It is important to recall here how random variables are defined in
Kolmogorov's setting (according to the measure theoretic approach).
See~\cite{Gudder-StatisticalMethods} for a detailed exposition.
A~random variable $f$ can be defined as a measurable function
$f:\Omega\longrightarrow\mathbb{R}$. In~this context, by~a
measurable function $f$, we mean a function satisfying that, for
every Borel subset $B$ of the real line (the \emph{borel~sets}
($B(\mathbb{R})$) are defined as the family of subsets of
$\mathbb{R}$ such that (a) it is closed under set theoretical
complements, (b) it is closed under denumerable unions, and~(c) it
includes all open intervals~\cite{ReedSimon}), we have that
$f^{-1}(B)\in\Sigma$ (i.e., the~pre-image of every Borel set $B$
under $f$ belongs to $\Sigma$, and~thus it has a definite
probability measure given by $\mu(f^{-1}(B))$).

Notice that a random variable $f$ defines an inverse map $f^{-1}$

\begin{subequations}\label{e:ClassicalPVM}
\begin{equation}\label{e:5a}
f^{-1}:B(\mathbb{R})\longrightarrow\Sigma
\end{equation}
\noindent satisfying
\begin{equation}
f^{-1}(\emptyset)=\emptyset
\end{equation}
\begin{equation}
f^{-1}(\mathbb{R})=\Omega
\end{equation}
\begin{equation}
f^{-1}(\bigcup_{j}B_{j})=\bigcup_{j}f^{-1}(B_{j})
\end{equation}
\noindent for any disjoint denumerable family of Borel sets
${B_{j}}$. {Denoting the complement of a set $X$ by $X^{c}$}, we
have that, for every Borel set $B$:
\begin{equation}\label{e:5e}
f^{-1}(B^{c})=(f^{-1}(B))^{c}.
\end{equation}

\end{subequations}

To illustrate ideas, let us consider a classical probabilistic
system. A~classical observable $H$ (such as the energy) will be a
function from the state space $\Gamma$ to the real numbers.
The~state of the system, given by a probability density $\varrho$
(i.e., $\varrho\geq 0$ and with Lebesgue integral
$\int_{\Gamma}\varrho(x)dx=1$)), will define a measure $\mu$ over
the measurable subsets of $\Gamma$ as follows. For~each subset
$S\in\Gamma$, define
\begin{equation}
\mu(S):=\int_{S}\varrho(x)dx.
\end{equation}

Measurable subsets of $\Gamma$ will be those for which the above
integral converges. The~function $\mu$ will obey Kolmogorov's
axioms, provided that we take $\Gamma=\Omega$ and $\Sigma$ as the
set of measurable subsets of $\Gamma$. The~above formula is
sufficient to compute the mean values and the probabilities of any
event of interest. Given an elementary testable proposition such as:
``the values of the observable $H$ lie in the interval $(a,b)$'',
the~real number $\mu( H^{-1}( (a,b) ) )$ gives us the probability
that this proposition is true. In~this sense, each observable of a
classical probabilistic system can be considered as a random
variable. This has to be a necessary condition for any admissible
classical state: a state must specify definite probabilities for
every elementary test that we may perform on the system. In~this
sense, each classical (probabilistic) state can be described by a
Kolmogorovian measure, with~the observables represented as random
variables.

At the same time, by~associating ``$\vee$'' with ``$\cup$'',
``$\wedge$'' with ``$\cap$'', ``$\neg$'' with ``$(\ldots)^{c}$''
(set theoretical complement), and ``$\leq$'' with ``$\subseteq$''
(set theoretical inclusion), we see that the Boolean structure
associated to measurable subsets is coincident with the distributive
character of classical logic. The~fact that the logic associated to
a classical system is Boolean (in the above operational sense), was
one of the main observations in~\cite{BvN}.

As we will see in the following sections, the~quantum formalism can
be considered as an extension of the classical one, provided that we
replace the measurable subsets of phase space with
$\mathcal{P}(\mathcal{H})$ (the lattice of projection operators on a
Hilbert space $\mathcal{H}$), the~measure $\mu$ by a quantum state
represented by a density operator, and~the classical random
variables by projection valued measures associated to self-adjoint
operators. As~a consequence, the~operational logic associated to a
quantum system will fail to be Boolean~\cite{BvN}, due to the
non-distributive character of $\mathcal{P}(\mathcal{H})$. The~set of
states of a quantum system will be convex too. However, the
geometrical shape will be very different from that of a classical
one, due to the non-Boolean character of the lattice of
events~involved.

\subsection{Cox's Approach}

Since the beginning of probability theory, there has been a school
of thought known as Bayesianism, which treats probabilities in a
different manner from the one discussed in the previous section.
For~the Bayesian approach, probabilities are not to be regarded as a
property of a system but~as a property of our knowledge about it.
This position is present as early as in the XIX century in one of
the milestones in the development of probability
theory~\cite{Laplace}. In~his work, Laplace proposed a way to assign
probabilities in~situations of ignorance that would eventually be
known as the``Laplace principle''. Later works would attempt to
formalize and give coherence to the Bayesian approach, as~for
example,~\cite{Jeffreys,deFinetti}. In~this section, we center our
attention on one of these attempts~\cite{CoxLibro,CoxPaper}, of R.
T.~Cox.

While attaining equivalent results to those of Kolmogorov, Cox's
approach is conceptually very different. In~the Kolmogorovian
approach, probabilities can be naturally interpreted (though not
necessarily) as relative frequencies in a sample space. On~the other
hand, in~the approach developed by Cox, probabilities are considered
as a measure of the degree of belief of a rational agent---which may
be a machine---on the truth of a proposition $x$, if~it is known
that proposition $y$ is true. In~this way, Cox intended to find a
set of rules for inferential reasoning that would be coherent with
classical logic and~that would reduce to it whenever all the
premises have definite truth~values.

To do this, he started with two very general axioms and presupposed
the calculus of classical propositions, which, as~is well known,
forms a Boolean lattice~\cite{Boole}. By~doing so, he derived
classical probability theory as an inferential calculus on Boolean
lattices. We sketch here the arguments presented in his
book~\cite{CoxLibro}. For~a more detailed exposition on the
deductions, the reader is referred
to~\cite{CoxPaper,CoxLibro,Jaynes,ReviewCox,Knuth-2004a,Knuth-2004b,Knuth-2005b}.
See~\cite{Terenin-Draper} for discussions on a rigorization of Cox's
method.

The two axioms used by Cox~\cite{CoxLibro} are

\begin{itemize}

\item C1---The probability of an inference on given evidence determines the probability of its contradiction from the same~evidence.

\item C2---The probability on a given evidence that both of two inferences are true is determined by their separate probabilities, one from the given evidence and the~other from this evidence with the additional assumption that the first inference is true.
\end{itemize}

A real valued function $\varphi$ representing the degree to which a
proposition $h$ (usually called \emph{hypothesis}) implies another
proposition $a$ is postulated. Thus, $\varphi(a|h)$ will represent
the degree of belief of an intelligent agent regarding how likely it
is that $a$ is true given that the agent knows that the hypothesis
$h$ is true.

Then, requiring the function $\varphi$ to be coherent with the
properties of the calculus of classical propositions, the agent
derives the rules for manipulating probabilities. Using axiom C2,
the~associativity of the conjunction ($a \wedge (b \wedge c) = (a
\wedge b) \wedge c)$), and~defining the function
$F[\varphi(a|h),\varphi(b|h)] \equiv \varphi(a \wedge b|h):
\mathbb{R}^2 \rightarrow \mathbb{R}$, the agent arrives at a
functional equation for $F(x,y)$:
\begin{equation}\label{eq:prod}
 F[x,F(y,z)] = F[F(x,y),z]
\end{equation}

Which, after~a rescaling and a proper definition of the probability
$P(a|h)$ in terms of $\varphi(a|h)$, leads to the well known
\emph{product rule} of probability theory:
\begin{equation}
 P(a \wedge b |h) = C P(a | h \wedge b) P(b|h)
\end{equation}

The definition of $P(a|h)$ in terms of $\varphi(a|h)$ is omitted,
as~one ultimately ends up using only the function $P(a|h)$ and never
$\varphi(a|h)$. In~an analogous manner, using axiom C1, the~law of
double negation ($\neg \neg a=a$), Morgan's law for disjunction
($\neg (a \vee b) = \neg a \wedge \neg b$), and defining the
function $f[P(a|h)] \equiv P(\neg a|h): \mathbb{R} \rightarrow
\mathbb{R}$, we arrive at the following functional equation for
$P(a|h)$
\begin{equation}\label{eq:neg}
 P(a|h)^r + P(\neg a|h)^r = 1
\end{equation}

With $r$ as an arbitrary constant. Although,~in principle, different
values of $r$ would give rise to different rules for the computation
of the probability of the negation of a proposition, as~taking
different values of $r$ account for a rescaling of $P(a|h)$, one
could as well call $P'(a|h) \equiv P(a|h)^r$ \emph{probability} and
work with this function instead of $P(a|h)$. For~simplicity, Cox
decided to take $r=1$ and to continue using $P(a|h)$.

Using Equations \eqref{eq:prod} and \eqref{eq:neg}, the~law of
double negation and Morgan's law for conjunction ($\neg (a \wedge b)
= \neg a \vee \neg b$), we arrive at the \emph{sum rule} of
probability theory:
\begin{equation}
 \label{eq:sum}
 P(a \vee b|h) = P(a|h) + P(b|h) - P(a \wedge b|h)
\end{equation}

As it turns out, $P(a|h)$---if suitably normalized---satisfies the
properties of an additive Kolmogorovian probability
(Equations~\eqref{e:1a}--\eqref{e:1c}).

Due to the importance of Cox's theorem to the foundations of
probability, it has been the target of thorough scrutiny by many
authors. Some have pointed out inconsistencies behind the implicit
assumptions made during its derivations, most notably the
assumptions behind the validity of Equation \eqref{eq:prod}. Since
then, there have been different proposals to save Cox's approach by
proving it using less restrictive axioms. In~\cite{CriticaCox},
a~discussion of the status of Cox proposal is presented as well as a
counterexample to it. For~a review on the subject, it is recommended
to consult~\cite{ReviewCox}.

Once the general properties of the function $P(a|h)$ are
established, the~next problem is to find a way to determine prior
probabilities (i.e., probabilities conditional only to the
hypothesis $h$). Although,~formally, one could assign prior
probabilities in any way coherent with the normalization used,
in~practical situations, one is compelled to assign them in a way
that they reflect the information contained in the hypothesis $h$.
A~possible way to do this is by using the MaxEnt
principle~\cite{Jaynes,Jaynes2}, which we will review shortly in the
next section. Other ways of assigning prior probabilities include
the Laplace principle~\cite{Jeffreys} and coherence with symmetry
transformations~\cite{Jaynes3}. Nevertheless, the~existence of a
general algorithm for assigning prior probabilities is still an
open~question.

\subsection{MaxEnt~Principle}\label{sec:MaxEnt}

This principle asserts that the assignment of the prior
probabilities from a hypothesis $h$ should be done by maximizing the
uncertainty associated with its distribution while respecting the
constrains imposed over them by $h$. Although~this may sound
paradoxical, by~maximizing the uncertainty of the prior
probabilities one avoids  assuming more information than that
strictly contained in $h$.

Taking Shannon's information measure $S[P] = - \sum_i P(a_i|h)
log[P(a_i|h)]$ as the measure of the uncertainty associated with the
distribution $P$, the~MaxEnt principle can be restated as: the prior
probabilities corresponding to the hypothesis $h$ are given by the
distribution that maximizes $S[P]$ subject to the constraints
imposed by $h$ on $P$. The~simplest example is given by the
hypothesis $h$ that imposes no constraints on $P$, in~which case $P$
results as the uniform distribution, and the MaxEnt principle
reduces to Laplace's. Different kinds of constraints result in
different prior probability distributions (PPD)~\cite{Jaynes2}.
In~\cite{TablaMaxEnt}, a table of some of the distributions obtained
in this way is presented. Although,~given a set of constraints, the
corresponding PPD can be readily computed, there is no general
method of translating a hypothesis $h$ into equivalent~constraints.

By means of the MaxEnt principle, classical and quantum equilibrium
statistical mechanics can be formulated on the basis of information
theory~\cite{Jaynes}. Assuming that the prior knowledge about the
system is given by $n$ expectation values of a collection of
physical quantities $R_j$, i.e.,~$\langle R_1 \rangle,\ldots,\langle
R_n \rangle$, then, the~most unbiased probability distribution
$\rho(x)$ is uniquely fixed by maximizing Shannon's logarithmic
entropy $S$ subject to the $n$ constraints
\begin{eqnarray}\label{e:conditionsmean} \langle
R_{i}\rangle=r_{i};\,\, {\rm for \,\,all\,\,} i.
\end{eqnarray}

\noindent In order to solve this problem, $n$ Lagrange multipliers
$\lambda_i$ must be~introduced.

In the process of employing the MaxEnt procedure, one discovers that
the information quantifier $S$ can be identified with the
equilibrium entropy of thermodynamics if our prior knowledge
$\langle R_1 \rangle,\ldots, \langle R_n \rangle$ refers to
extensive quantities~\cite{Jaynes}. $S(maximal)$, once determined,
yields complete thermodynamical information with respect to the
system of interest~\cite{Jaynes}. The~MaxEnt probability
distribution function (PDF), associated to
Boltzmann--Gibbs--Shannon's logarithmic entropy $S$, is given
by~\cite{Jaynes}
\begin{equation} \label{z1}
\rho_{max}=\exp{(-\lambda_{0}\mathbf{1}-\lambda_{1}R_{1}-\cdots-\lambda_{n}R_{n})},
\end{equation}
where the $\lambda$'s are Lagrange multipliers guaranteeing that
\begin{equation}  \label{z2}
r_{i}=-\frac{\partial}{\partial\lambda_{i}}\ln Z,
\end{equation}
while the partition function reads
\begin{equation}  \label{z3}
Z(\lambda_{1}\cdots\lambda_{n})=\sum_{i}\exp^{-\lambda_{1}R_{1}(x_{i})-\cdots-\lambda_{n}R_{n}(x_{i})},
\end{equation}
and the  normalization condition
\begin{equation}  \label{z4}
\lambda_{0}=\ln Z.
\end{equation}

In a quantum setting, the~$R$'s are operators on a Hilbert space
$\mathcal{H}$, while $\rho$ is a density matrix (operator). The~sum
in the partition function must be replaced by a trace, and Shannon's
entropy must be replaced by von~Neumann's.

\section{The Formalism Of~QM}\label{s:TheFormalismOfQM}

In this Section, we discuss some specific features of the quantum
formalism~\cite{mikloredeilibro,vadar68,vadar70,vN} that are
relevant for the problem of~QP.

\subsection{Elementary Measurements And Projection~Operators}\label{s:ElementaryTests}

In QM, observable physical magnitudes are represented by compact
self-adjoint operators in a Hilbert space $\mathcal{H}$ (we denote
this set by $\mathcal{A}$). Due to the spectral decomposition
theorem~\cite{ReedSimon,vN}, a~key role is played by the notion of
\textit{projection valued measure} (PVM): the set of PVMs can be put
in a bijective correspondence with the set $\mathcal{A}$ of self
adjoint operators of $\mathcal{H}$. Intuitively speaking, a~PVM is a
map that assigns a projection operator to each interval of the real
line. In~this sense, projection operators are the building blocks
out of which any observable can be built. It is important to recall
that projection operators have a very clear operational meaning:
they represent elementary empirical tests with only two outputs
(zero and one, or~YES and NO). In~a formal way, a~PVM is a map $M$
defined over the Borel sets (see Section~\ref{s:RandomVariables}) as
follows

\begin{subequations}\label{e:PVM}
\begin{equation}\label{e:16a}
M: B(\mathbb{R})\rightarrow\mathcal{P}(\mathcal{H}),
\end{equation}
\noindent satisfying
\begin{equation}
M(\emptyset)=\mathbf{0}\,\,\,
(\mathbf{0}:=\mbox{null}\,\mbox{subspace})
\end{equation}
\begin{equation}
M(\mathbb{R})=\mathbf{1}
\end{equation}
\begin{equation}
M(\cup_{j}(B_{j}))=\sum_{j}M(B_{j}),\,\,
\end{equation}
\noindent for any disjoint denumerable family ${B_{j}}$.
\begin{equation}\label{e:16e}
M(B^{c})=\mathbf{1}-M(B)=(M(B))^{\bot}
\end{equation}

\end{subequations}

As we will see in the following Section, a~PVM is the natural
generalization of the notion of random variable to the non-Boolean
setting. In~order to realize why this is so, it is important to
compare Equations \eqref{e:5a}--\eqref{e:5e} and
\eqref{e:16a}--\eqref{e:16e}. It is also important to remark that
the set of projections in the image of a PVM are always orthogonal:
this implies that this set can always be endowed with a Boolean
lattice structure. This allows us to associate, to each complete
observable, a~particular empirical context represented by a Boolean
algebra of events. Thus, in~this sense, complete observables are
always referred to a particular~context.

Fixing an element $A\in\mathcal{A}$, the~intended interpretation of
the associated PVM ($M_{A}(\ldots)$), evaluated in an interval
$(a,b)$ (i.e., $M_{A}((a,b))=P_{(a,b)}$) is: ``the value of $A$ lies
between the interval $(a,b)$''. In~this sense, projection operators
represent elementary tests or propositions in QM. In~other words,
they can be considered as the simplest quantum mechanical
observables. As~we reviewed in Appendix~\ref{s:LatticeTheory},
projection operators can be endowed with a lattice structure
and,~thus, also elementary tests. This lattice was called ``Quantum
Logic'' by Birkhoff and von Neumann~\cite{BvN}. We refer to it as
the \emph{von Neumann-lattice
($\mathcal{P}(\mathcal{H})$})~\cite{mikloredeilibro}. As shown
in~\cite{BvN}, an~analogous treatment can be done for classical
systems. As~we have seen in Section~\ref{s:RandomVariables},
propositions associated to a classical system are endowed with a
natural Boolean~structure.

During the thirties, von Neumann and collaborators continued
studying formal developments related to the quantum formalism. One
of the results of this investigation was the development of the
theory of \emph{rings of operators} (better known as \emph{von
Neumann algebras}
\cite{mikloredeilibro,RingsOfOperatorsI,RingsOfOperatorsII,RingsOfOperatorsIII,RingsOfOperatorsIV}),
as an attempt of generalizing certain algebraic properties of Jordan
algebras~\cite{vN-Hilbert-Nordheim}. The~subsequent study of von
Neumann algebras showed that they are closely related to lattice
theory. Murray and von Neumann provided a classification of factors
(von Neumann algebras whose center is formed by the multiples of the
identity) using orthomodular lattices
in~\cite{RingsOfOperatorsI,RingsOfOperatorsII,RingsOfOperatorsIII,RingsOfOperatorsIV}.
On the other hand, lattice theory is deeply connected to projective
geometry~\cite{ProjectiveGeometries}, and~one of the major
discoveries of von Neumann was that of \emph{continuous geometries},
which do not possess ``points'' (or ``atoms'') and are related to
type II factors. Far from being a mere mathematical curiosity, type
II factors found applications in statistical mechanics and type III
factors play a key role in the axiomatic approach to Quantum Field
Theory (QFT)~\cite{Redei-Summers2006,mikloredeilibro}.

The quantum logical approach of Birkhoff and von Neumann was
continued by other
researchers~\cite{Gudder-StatisticalMethods,aertsdaub1,jauch,piron,mackey-book,vadar68,vadar70}
(see~\cite{dallachiaragiuntinilibro,HandbookofQL,kalm83} for
complete expositions). One of the key results of this approach is
the \emph{representation theorem} of C. Piron~\cite{piron}. He
showed that any propositional system can be coordinatized in a
generalized Hilbert space. A~later result by Sol\`{e}r showed that,
by adding extra assumptions, it can only be a Hilbert space over the
fields of the real numbers, complex numbers, or
quaternions~\cite{Soler-1995}.

\subsection{Quantum States And Quantum~Probabilities}\label{s:QuantumProbabilities}

In this Section we discuss QP. We do this by reviewing the usual
approach, in~which Kolmogorov's axioms are extended to non-Boolean
lattices (or algebras)~\cite{Redei-Summers2006}.

As we have seen in Section~\ref{s:ElementaryTests}, elementary tests
in QM are represented by closed subspaces of a Hilbert space. These
subspaces form an orthomodular atomic lattice
$\mathcal{P}(\mathcal{H})$. In~order to assign probabilities to
these elementary tests or processes, many texts proceed by
postulating axioms that are similar to those of
Kolmogorov~\cite{mikloredeilibro,vN,belcas81}. The~Boolean
$\Sigma$-algebra appearing in Kolmogorov's axioms
(Equations~\eqref{e:1a}--\eqref{e:1c}) is replaced by
$\mathcal{P}(\mathcal{H})$, and~a measure $s$ is defined as follows:

\begin{subequations}\label{e:nonkolmogorov}
\begin{equation}\label{e:17a}
s:\mathcal{P}(\mathcal{H})\rightarrow [0;1]
\end{equation}
\noindent such that:
\begin{equation}\label{e:Qprobability1}
s(\textbf{1})=1 \,\, (\textbf{1}:=\mathcal{H})
\end{equation}
\noindent and, for~a denumerable and pairwise orthogonal family of
projections ${P_{j}}$,
\begin{equation}\label{e:17c}
s(\sum_{j}P_{j})=\sum_{j}s(P_{j}).
\end{equation}
\end{subequations}

In this way, a~real number between $0$ and $1$ is assigned to any
elementary test. Despite  the similarity with Kolmogorov's axioms,
the~probabilities defined above are very different, due to the
non-Boolean character of the lattice involved. Gleason's
theorem~\cite{Gleason,Gleason-Dvurechenski-2009} asserts that if
$dim(\mathcal{H})\geq 3$, any measure $s$ satisfying
\eqref{e:17a}--\eqref{e:17c} can be put in correspondence with a
trace class operator (of trace one) $\rho_{s}$:
\begin{equation}\label{e:bornrule2}
s(P):=\mbox{tr}(\rho_{s} P)
\end{equation}

\noindent for every orthogonal projection $P$. On~the other hand,
using Equation \eqref{e:bornrule2}, any trace class operator of
trace one defines a measure as in \eqref{e:17a}--\eqref{e:17c}, and
thus the correspondence is one to one for $dim(\mathcal{H})\geq 3$
(something that is not true for the two dimensional case). In~this
way, Equations \eqref{e:17a}--\eqref{e:17c} define the usual
probabilities of QM and constitute a natural generalization of
Kolmogorov's axioms to the quantum~case.

The set $\mathcal{C}(\mathcal{H})$ of all possible measures
satisfying Equations \eqref{e:17a}--\eqref{e:17c} is indeed convex,
as~in the classical case. However, these sets are very different.
As~an example, let us compare a classical bit (to fix ideas, think
about the possible probabilistic states of a coin) and a qubit (a
quantum system represented by a two-dimensional model). While the
state space of the first one is a line segment, it is well known
that the state space of the second is homeomorphic to a
sphere~\cite{BEN}. For more discussion about the convex set of
quantum states, see~\cite{Holik-Zuberman,HolikPRA-2011}.

A state satisfying Equations \eqref{e:17a}--\eqref{e:17c} will yield
a Kolmogorovian probability when restricted to a maximal Boolean
subalgebra of $\mathcal{P}(\mathcal{H})$. In~this way, a~quantum
state can be considered as a coherent pasting of different
Kolmogorovian measures. This has a natural physical interpretation
as follows. Each empirical setup will define a maximal Boolean
algebra. The~fact that quantum states yield the correct observed
frequencies (via the Born rule), allows defining consistent
Kolmogorovian probabilities on each Boolean setting. However, doing
statistics on repeated measurements in identical preparations using
a single empirical setup, is not sufficient to determine a quantum
state completely.

For~a general state, it will be mandatory to perform measurements in
different and complementary (in Bohr's sense) empirical setups.
Notice that there are many ways in which one could define a family
of Kolmogorovian probabilities in $\mathcal{P}(\mathcal{H})$.
However, the probabilities defined by a quantum state (or
equivalently, by~Equations \eqref{e:17a}--\eqref{e:17c}), have a
very particular mathematical form. The~existence of uncertainty
relations between non-compatible observables~\cite{Holik-MachZender}
is nothing but an expression of this fact. The fact that a quantum
state $\mu$ can be considered as a coherent collection of
Kolmogorovian measures can be summarized in a diagram as follows.
{If $\Sigma$ is an arbitrary Boolean subalgebra of
$\mathcal{P}(\mathcal{H})$, let $\mu$ be a quantum state and
$\mu_{\Sigma}$ the restriction of $\mu$ to $\Sigma$. Then, for every
Boolean subalgebra $\Sigma$, we have the following commutative
diagram:
\[
\xymatrix{
\Sigma\ar@{^{(}->}[r]\ar[dr]^-{\mu_{\Sigma}}&\mathcal{P}(\mathcal{H})\ar[d]^{\mu}\\
& [0,1]}
\]

\noindent The fact that there exists a global quantum state $\mu$
that makes the above diagram commute for every Boolean subalgebra,
is a quite remarkable fact about the quantum formalism. Notice that,
given that the intersection of Boolean subalgebras may be
non-trivial (see examples in the next Section), the~probability
assignments must satisfy certain compatibility conditions. Thus,
even if an event $x$ belongs to two different measurement contexts,
the quantum state assigns to it the same probability. That is,
the~probability is assigned independently of the context to which it
belongs. This is known as the \emph{no-signal} condition, which will
not necessarily hold outside physics (for example, in~cognition
experiments).}

The above generalization also includes quantum observables in a
natural way. Indeed, by~appealing to the spectral decomposition
theorem, there is a one-to-one correspondence between quantum
observables represented by self-adjoint operators and PVM's. Then,
these notions are interchangeable. However, a quick look to
Equations \eqref{e:16a}--\eqref{e:16e} reveals that PMS are very
similar to classical random variables: while classical random
variables map Borel sets into the Boolean lattice of measurable
sets, PVMs map Borel sets into the non-Boolean lattice
$\mathcal{P}(\mathcal{H})$. Thus, quantum observables can be
reasonably interpreted as non-Kolmogorovian random~variables.

We  mentioned above that \eqref{e:1a}--\eqref{e:1c} and
\eqref{e:17a}--\eqref{e:17c} are not equivalent probability
theories. For~example, Equation~\eqref{e:SumRule} is no longer valid
in QM. Indeed, for~suitably chosen $s$ and quantum events $A$ and
$B$, we have
\begin{equation}\label{e:QuantumInequality}
s(A)+s(B)\leq s(A\vee B)
\end{equation}

The above inequality should be compared with the classical one,
given by \eqref{e:SumRuleInequality}. {As an example, consider a two
dimensional quantum system (in current jargon: a qubit), the~events
$A= |\uparrow_{z}\rangle\langle\uparrow_{z}|$ (spin up in direction
$\hat{z}$) and $B= |\downarrow_{x}\rangle\langle\downarrow_{x}|$
(spin down in direction $\hat{x}$), and~the state
$|\psi\rangle=\frac{1}{\sqrt{2}}(|\uparrow\rangle_{z}+|\downarrow\rangle_{z})$
(``cat state'' in the basis $\hat{z}$, which is the same as ``spin
up in direction $\hat{x}$''). Thus, using some simple math, we
obtain that $p_{|\psi\rangle}(A)=
\mbox{tr}(|\psi\rangle\langle\psi||\uparrow_{z}\rangle\langle\uparrow_{z}|)=\frac{1}{2}$,
and $p_{|\psi\rangle}(B)=
\mbox{tr}(|\psi\rangle\langle\psi||\downarrow_{x}\rangle\langle\downarrow_{x}|)=0$.
On the other hand, we have that $A\vee B$ is the linear span of
$|\uparrow_{z}\rangle$ and $|\downarrow_{x}\rangle$, and~then,
$A\vee B=I$. Since $p_{|\psi\rangle}(I)=
\mbox{tr}(|\psi\rangle\langle\psi|I)=1$, we have that
$p_{|\psi\rangle}(A)+p_{|\psi\rangle}(B)=\frac{1}{2}+0<1=p_{|\psi\rangle}(A\vee
B)$.}

The probability theory defined by \eqref{e:17a}--\eqref{e:17c} can
also be considered as a non-commutative generalization of classical
probability theory in the following sense: while when in an
arbitrary statistical theory, a~state will be a normalized measure
over a suitable $C^{\ast}$-algebra, the~classical case is recovered
when the algebra is \emph{commutative}
\cite{Gudder-StatisticalMethods,Redei-Summers2006}. We end  this
Section by noting that some technical complications appear when one
attempts to define a quantum conditional probability in the
non-commutative setting. For~a complete discussion about these
matters and a comparison between classical and quantum
probabilities,
see~\cite{Gudder-StatisticalMethods,Redei-Summers2006}.

\subsection{Some~Examples}

In order to understand better the mathematical structure (and the
physical interpretation) underlying quantum probabilities, we
discuss here some examples in detail. We relate the event structures
associated to physical systems with different notions of lattice
theory. We do this by enumerating different examples that are
relevant for the discussion presented in this work. The~reader
unfamiliar with lattice theory can consult
Appendix~\ref{s:LatticeTheory}.

\begin{enumerate}
\item \textbf{Finite Probability model: a dice.}
Consider the throw of a dice. The~possible outcomes are given by
$\Omega=\{1,2,3,4,5,6\}$. A~probabilistic state of the dice is
determined by assigning real numbers $p_{i}$, $i=1,...,6$, to~each
element of $\Omega$. If~the dice is not loaded, then
$p_{i}=\frac{1}{6}$ for all $i$; however,~a realistic dice will not
satisfy this. An~event will be represented by a subset of $\Omega$.
As examples, consider the event ``the outcome is even'' or ``the
outcome is greater than 2''. These are represented by $\{2,4,6\}$
and $\{3,4,5,6\}$, respectively. All possible subsets of $\Omega$
form a Boolean lattice (see Apendix \ref{s:LatticeTheory}),
with~regard to the set-theoretical operations: ``$\cup$''
(interpreted as ``$\vee$''), ``$\cap$'' (interpreted as
``$\wedge$''), and~the set theoretical complement (interpreted as
``$\neg$''). The~example of a $\sigma-$algebra associated to a
measurable space $(\Omega,\Sigma,\mu)$ works in a similar~way.

\item \textbf{Hilbert lattice:} As discussed above, the~events associated
to quantum systems can be put in one to one correspondence with an
orthomodular lattice: the one formed by the set of closed subspaces
of a Hilbert space $\mathcal{H}$. They can be endowed with a lattice
structure as follows~\cite{mikloredeilibro}. The~operation``$\vee$''
is taken as the closure of the direct sum ``$\oplus$'' of subspaces,
``$\wedge$'' as the intersection ``$\cap$'', and~``$\neg$'' as the
orthogonal complement ``$\bot$'', $\textbf{0}=\vec{0}$,
$\textbf{1}=\mathcal{H}$, and~we denote by
$\mathcal{P}(\mathcal{H})$
 the set of closed subspaces. The~order ``$\leq$'' is defined by
subspace inclusion: we say that $\mathbb{S}\leq\mathbb{T}$, whenever
$\mathbb{S}\subseteq\mathbb{T}$. The~subspaces $\textbf{0}$ and
$\textbf{1}$ play the role of the bottom and top elements of the
lattice, since, for any subspace $\mathbb{S}$, we have
$\textbf{0}\leq\mathbb{S}\leq\textbf{1}$.

Then, the~algebraic structure $(\mathcal{P}(\mathcal{H}),\ \cap,\
\oplus,\ \neg,\ 0,\ 1)$ will be a complete bounded orthomodular
lattice (which we denote simply by $\mathcal{P}(\mathcal{H})$). It
is complete because~the intersections and (the closure of) sums of
arbitrary families of closed subspaces yields closed subspaces. It
is bounded due to the existence of the top and bottom elements. It
is orthomodular, because for any pair of subspaces $\mathbb{S}$, and
$\mathbb{T}$, whenever we have $\mathbb{S}\leq\mathbb{T}$, then
$\mathbb{S}\vee((\mathbb{S})^{\bot}\wedge\mathbb{T})=\mathbb{T}$
(see Appendix \ref{s:LatticeTheory} for more details).

As closed subspaces are in one to one correspondence with projection
operators, we take $\mathcal{P}(\mathcal{H})$ to be the lattice of
closed subspaces or the lattice of projections interchangeably. One
of the most important features of $\mathcal{P}(\mathcal{H})$ is that
the distributive law \eqref{e:DistributiveLaw} does not hold (see
Appendix \ref{s:LatticeTheory}). $\mathcal{P}(\mathcal{H})$ is
modular if $\mathcal{H}$ is finite dimensional. If~$\mathcal{H}$ is
infinite dimensional, then $\mathcal{P}(\mathcal{H})$ is always
orthomodular. Gleason's theorem (mentioned in the previous section)
grants that, for~$dim(\mathcal{H})\geq 3$, quantum states can be
considered as measures over the lattice of closed subspaces. This is
a remarkable fact, since it implies that quantum probabilities are
described by a very specific mathematical~framework.

Any measurement context can be represented by an orthogonal basis of
$\mathcal{H}$. It is easy to check that, by applying the lattice
operations defined above to a fixed basis, we obtain a Boolean
algebra. The~cases of Hilbert spaces of dimension $2$ and $3$ are
easy to check; however,~this is true in general. It turns out that
the whole lattice of subspaces can be described as a family of
intertwined Boolean algebras (more about this below). For~more
discussion regarding the notion of ``intertwined contexts'',
see~\cite{Svozil-Intertwined}.

The following table summarizes the main differences between Quantum
and Kolmogorovian probabilities:

\vspace{6pt}
\begin{tabular}{ c c c}
    & \textbf{Kolmogorov Probability} & \textbf{Quantum Probability} \\
\textbf{Lattice:} & $\Sigma$ &
$\mathcal{P}(\mathcal{H})$\\
         &         (Boolean-algebra)   & (orthomodular, non-Boolean)\\
\textbf{States:} & Measures over $\Sigma$ & Measures over $\mathcal{P}(\mathcal{H})$\\
\textbf{Events:} & Subsets of $\Omega$  & Closed subspaces of $\mathcal{H}$\\
\end{tabular}
\vspace{6pt}

There is a geometrical underlying quantum probability: the one
dimensional subspaces of a Hilbert space form a \emph{projective
geometry}. The~higher dimensional subspaces are elements of the
\emph{projective lattice} associated with this geometry
(see~\cite{vadar68,vadar70}).

\item \textbf{Firefly Model:}
The firefly model~\cite{svozillibro}
is used in quantum logic to show an example of a system that is not
a full quantum model but~has certain features that serve to
illustrate what happens with quantum systems. It consists of a
firefly that is freed inside a box. We are asked to perform an
experiment to detect the location of the firefly but~with
constrains. We are only allowed to look at two different faces of
the box (and we can only choose one on each run of the experiment).
The~first one is to measure face $C_{1}$ with~three possible
outcomes: the ``firefly is detected on the left'' (l), on~the right
(r), and~``no-signal'' (n) (which means that the light of the
firefly was off).

The~second possibility is to measure on face $C_{2}$, with~the
possibilities ``firefly is on front'' (f), ``firefly is in the
bottom'' (b), and~``no-signal''. Notice that the ``no-signal''
outcome (n) is present in both experiments---this will be important
soon. These constraints are, of~course, silly, given that we can
always look at every place in the box and detect the exact location
of the firefly. However, they are thought off as an artificial
measurement procedure that resembles what happens with quantum
systems. If~we choose context $C_1$, we can check whether the
firefly is on the left, the~right, or~no signal. If~we choose
context $C_{2}$, we can check whether it is front, bottom,
or~no-signal. However, we cannot check both things in the same
experiment, as~happens with the position and momentum of a quantum
system.

If we choose to measure in the face $C_1$, the~three outcomes form
an outcome set $\Omega_{1}=\{(l),(r),(n)\}$. This gives rise to a
Boolean algebra $\Sigma_{1}$, formed by all possible subsets of
$\Omega_1$. Each of these subsets represents an event, such as ``the
firefly is not detected on the right'' (which is represented by the
set $\{(l),(n)\}$), and~so on. A~probabilistic state of the firefly
---a throw in which we do not know the outcome a priori---will give a
classical probability space $(\Omega_{1},\Sigma_{1},\mu_{1})$.
Similarly, we have a probability space
$(\Omega_{2},\Sigma_{2},\mu_{2})$ for the second option $C_2$, where
$\Omega_2 = \{(t),(b),(n)\}$. Notice that
$\Omega_1\cap\Omega_2=\{\mathbf{0},(n),(n)',\mathbf{1}\}$. Since the
event $(n)$ belongs to both contexts of measurement, for~the sake of
consistency, we must impose $\mu_{1}((n))=\mu_{2}(n)$.

Now, let $(n')=\{(l),(r)\}$, $(l')=\{(n),(r)\}$ and
$(r')=\{(n),(l)\}$ (i.e., the~set of theoretical complements of
$(n)$, $(l)$ and $(r)$, respectively). The~Hasse diagram of
$\Sigma_{1}$ is then given~by

\vspace{6pt}
  \begin{tikzpicture}
  \node (max) at (0,4) {$\mathbf{1}$};

  \node (a) at (-2,2) {$(l)'$};

  \node (b) at (0,2) {$(r)'$};

  \node (c) at (2,2) {$(n)'$};

  \node (f) at (-2,0) {$(l)$};

  \node (g) at (0,0) {$(r)$};

  \node (h) at (2,0) {$(n)$};

  \node (min) at (0,-2) {$\mathbf{0}$};

  \draw (min) -- (f) -- (b) -- (max);

  \draw (min) -- (g) -- (a) -- (max);

  \draw (min) -- (h) -- (a) -- (max);

  \draw (h) -- (b) -- (max);

  \draw (f) -- (c) -- (max);

  \draw (g) -- (c) -- (max);

\end{tikzpicture}
\vspace{6pt}

In the above diagram, a~line joining two elements $x$ and $y$, means
that $x\leq y$ (i.e., the~partial order is represented by the lines
connecting the different elements). Thus, for~example, $(l)\leq(r)'$
(which is equivalent to $\{(l)\}\subseteq\{(n),(l)\}$). The~join of
two elements is the least element that lies above both of them (with
regards to the partial order). The~conjunction is the greatest
element that lies below both. Thus, for~example, $(l)\vee(r)=(n)'$
(which means $\{(l)\}\cup\{(r)\}=(\{(n)\})^{c}$) and
$(l)\wedge(r)=\mathbf{0}$ (which means
$\{(l)\}\cap\{(r)\}=\emptyset$). A~similar convention holds for the
rest of the diagrams~below.

Similarly, the~Hasse diagram of $\Sigma_{2}$ is then given~by

\vspace{6pt}
  \begin{tikzpicture}
  \node (max) at (0,4) {$\mathbf{1}$};

  \node (a) at (-2,2) {$(n)'$};

  \node (b) at (0,2) {$(f)'$};

  \node (c) at (2,2) {$(b)'$};

  \node (f) at (-2,0) {$(n)$};

  \node (g) at (0,0) {$(f)$};

  \node (h) at (2,0) {$(b)$};

  \node (min) at (0,-2) {$\mathbf{0}$};

  \draw (min) -- (f) -- (b) -- (max);

  \draw (min) -- (g) -- (a) -- (max);

  \draw (min) -- (h) -- (a) -- (max);

  \draw (h) -- (b) -- (max);

  \draw (f) -- (c) -- (max);

  \draw (g) -- (c) -- (max);

\end{tikzpicture}
\vspace{6pt}

A direct check shows that $\Sigma_{1}$ and $\Sigma_{2}$ are Boolean
algebras (also, Boolean lattices---see
Appendix~\ref{s:LatticeTheory}). Now, we can join all possible
events together, taking into account that
$\Sigma_{1}\cap\Sigma_{2}=\{\mathbf{0},(n),(n)',\mathbf{1}\}$. We
obtain the following Hasse diagram:

\vspace{6pt}
  \begin{tikzpicture}
  \node (max) at (0,4) {$\mathbf{1}$};

  \node (a) at (-4,2) {$(l)'$};

  \node (b) at (-2,2) {$(r)'$};

  \node (c) at (0,2) {$(n)'$};

  \node (d) at (2,2) {$(f)'$};

  \node (e) at (4,2) {$(b)'$};

  \node (f) at (-4,0) {$(l)$};

  \node (g) at (-2,0) {$(r)$};

  \node (h) at (0,0) {$(n)$};

  \node (i) at (2,0) {$(f)$};

  \node (j) at (4,0) {$(b)$};

  \node (min) at (0,-2) {$\mathbf{0}$};

  \draw (min) -- (f) -- (b) -- (max);

  \draw (min) -- (g) -- (a) -- (max);

  \draw (min) -- (h) -- (a) -- (max);

  \draw (min) -- (i) -- (e) -- (max);

  \draw (min) -- (j) -- (d) -- (max);

  \draw (h) -- (e) -- (max);

  \draw (h) -- (d) -- (max);

  \draw (h) -- (b) -- (max);

  \draw (h) -- (d) -- (max);

  \draw (f) -- (c) -- (max);

  \draw (i) -- (c) -- (max);

  \draw (j) -- (c) -- (max);

  \draw (g) -- (c) -- (max);

\end{tikzpicture}
\vspace{6pt}

The above diagram defines a lattice $\mathcal{L}$, which---like
$\Sigma_{1}$ and $\Sigma_{2}$---is non-distributive (and thus,
non-Boolean). The~lattice join of two given elements is the least
element that lies above them, and~the conjunction is the greatest
element from below. The~reader can check non-distributivity by
inspection.

The~Boolean algebras $\Sigma_{1}$ and $\Sigma_{2}$ are sublattices
of $\mathcal{L}$. It is very important to remark that they contain
an element in common: $\mathcal{L}$ can be seen a pasting of
$\Sigma_{1}$ and $\Sigma_{2}$. In~other words, $\mathcal{L}$ is
formed from two Boolean subalgebras that are \emph{intertwined}.
The~associated lattices of fully quantum models are just like that:
they are formed by a collection of intertwined Boolean
subalgebras---one for each context. The~difference between the
lattice of the firefly and the lattice of a three-dimensional
quantum system is that there are infinitely many contexts for the
latter, and thus the intertwining---for $dim(\mathcal{H})\geq
3$---is much more complicated. This intricate algebraic structure
associated to quantum systems lies at the core of the celebrated
Kochen--Specker theorem~\cite{KS} (which we discuss below).

\item \textbf{The lattice of Q-bit:}
Given the incredible advances of quantum information theory in
recent decades, the~reader may wonder what the lattice of a q-bit
looks like. It is the simplest quantum model conceivable. Suppose
then that we are given a spin $\frac{1}{2}$ system. As~is well
known, the~set of all possible states of a qubit is isomorphic to a
sphere, namely, the~\emph{Bloch sphere} \cite{BEN}. \emph{Each pure
state of a qubit corresponds to a one dimensional subspace of a two
dimensional complex Hilbert space, and can be represented as a point
in the surface of the Bloch sphere}.

A~one dimensional subspace is called a \emph{ray}. The~different
sets of objective properties, which are of the form ``the particle
has spin $\uparrow$ (or $\downarrow$) in direction $\hat{n}$'', are
represented by those rays (or, equivalently, by~points on the
surface of the sphere). Notice that each direction in space
$\hat{n}$ defines two rays in the Hilbert space (represented by the
projection operators
$P^{\uparrow}_{\hat{n}}=|\uparrow\rangle\langle\uparrow|_{\hat{n}}$
and
$P^{\downarrow}_{\hat{n}}=|\downarrow\rangle\langle\downarrow|_{\hat{n}}$).

The subspaces associated to $P^{\uparrow}_{\hat{n}}$ and
$P^{\downarrow}_{\hat{n}}$ are orthogonal: this means, literally,
that we must imagine them as orthogonal lines in the Hilbert space.
As there are infinitely many directions in space, there are
infinitely many such pairs of orthogonal events. All these events
will be included in the lattice of a~qubit.

In addition to all possible rays (associated to one dimensional
subspaces), we also have two distinguished subspaces, represented by
the events $\textbf{0}$ (the null subspace of the Hilbert space),
and~$\textbf{1}$ (the maximal subspace, which equals $\mathcal{H}$).
Each one dimensional subspace contains $\textbf{0}$ as a subspace,
and is contained in $\textbf{1}$. If~we chose a direction in space
$\hat{n}$,~consider the set
$\mathcal{B}_{\hat{n}}=\{\textbf{0},P^{\uparrow}_{\hat{n}},P^{\downarrow}_{\hat{n}},\textbf{1}\}$,
and consider the above defined lattice operations for subspaces, we
obtain a two-element Boolean algebra. All contexts of a qubit are of
this form: each measurement direction $\hat{n}$ in space defines a
two-elements Boolean algebra $\mathcal{B}_{\hat{n}}$. The~Hasse
diagram of a context represented by $\mathcal{B}_{\hat{n}}$ is then
given~by

\vspace{6pt}
\begin{tikzpicture}

  \node (max) at (0,1) {$\mathbf{1}$};

  \node (a) at (1,0) {$P^{\downarrow}_{\hat{n}}$};

  \node (b) at (-1,0) {$P^{\uparrow}_{\hat{n}}$};

  \node (min) at (0,-1) {$\mathbf{0}$};

  \draw (min) -- (a) -- (max);

  \draw (min) -- (b) -- (max);

\end{tikzpicture}
\vspace{6pt}

Thus, the~Hasse diagram of a q-bit will have the form:

\vspace{6pt}
\begin{tikzpicture}

  \node (max) at (0,1) {$\mathbf{1}$};

  \node (f) at (3,0) {$\cdots$};

  \node (a) at (2,0) {$P^{\downarrow}_{\hat{n}}$};

  \node (b) at (1,0) {$P^{\uparrow}_{\hat{n}}$};

  \node (e) at (0,0) {$\cdots$};

  \node (c) at (-1,0) {$P^{\downarrow}_{\hat{n}'}$};

  \node (d) at (-2,0) {$P^{\uparrow}_{\hat{n}'}$};

  \node (g) at (-3,0) {$\cdots$};

  \node (min) at (0,-1) {$\mathbf{0}$};

  \draw (min) -- (a) -- (max);

  \draw (min) -- (b) -- (max);

  \draw (min) -- (c) -- (max);

  \draw (min) -- (d) -- (max);

\end{tikzpicture}
\vspace{6pt}

\noindent where $\hat{n}$, $\hat{n}'$, etc., define different
directions in space. The~dots represent the infinitely many other
Boolan algebras associated with all possible directions in space.
Again, we obtain a lattice, which is non-distributive. In~this
example,
$\mathcal{B}_{\hat{n}}\cap\mathcal{B}_{\hat{n}'}=\{\mathbf{0},\mathbf{1}\}$
whenever $\hat{n}$ and $\hat{n}'$ define different directions. Thus,
only the top and bottom elements are shared by the Boolean
subalgebras. Thus, this example is degenerated, since there is no
(non-trivial) intertwining between the different Boolean algebras
associated to the measurement contexts. In~the following example, we
will consider a higher dimensional example, for~which the
intertwining is highly~non-trivial.

\item \textbf{Kochen--Specker theorem (in a four dimensional model):}
A nice example of how the different contexts of a quantum system are
intertwined was presented in~\cite{Cabello-KS-1996} (of course, for
the~original version of the Kochen--Specker theorem, the~reader is
referred to~\cite{KS}). Given a four-dimensional quantum system,
each measurement context has four possible outcomes. Each one of
them is mathematically represented by a one dimensional subspace of
a four-dimensional Hilbert space.

Each one dimensional subspace is generated by a vector $\hat{v}$.
Let us then represent the outcome given by the vector $\hat{v}$ by
the projection operator $P_{\hat{v}}$ (which is the projection
operator that projects into the subspace generated by $\hat{v}$).
Then, each measurement context is represented by four projection
operators, say $P_{\hat{v}_{1}}$ $P_{\hat{v}_{2}}$,
$P_{\hat{v}_{3}}$ and $P_{\hat{v}_{4}}$. These are all orthogonal,
because~they represent mutually exclusive outcomes (you cannot have,
in~the same measurement context, two different outputs at the same
time). As~in the qubit case, by~using the Hilbert lattice
operations, these projections generate a Boolean algebra with four
atoms (that has $2^{4}$ elements).

Thus, each measurement context has an associated Boolean algebra
with sixteen elements. When a quantum state is prepared,
the~probabilities assigned to each context can be thought off as
Kolmogorovian. It also happens that
$P_{\hat{v}_{1}}+P_{\hat{v}_{2}}+P_{\hat{v}_{3}}+P_{\hat{v}_{4}}=\mathbf{1}$
(which, using the Hilbert lattice operations, reads
$\bigvee_{i}P_{\hat{v}_{i}}=\mathbf{1}$). Now, consider a family of
nine different measurement contexts, giving place to the equations:
\begin{align}\label{e:CabelloKS}
P_{0,0,0,1}+P_{0,0,1,0}+P_{1,1,0,0}+P_{1,-1,0,0} &
=\hat{1}\nonumber\\\nonumber
P_{0,0,0,1}+P_{0,1,0,0}+P_{1,0,1,0}+P_{1,0,-1,0} &
=\hat{1}\\\nonumber
P_{1,-1,1,-1}+P_{1,-1,-1,1}+P_{1,1,0,0}+P_{0,0,1,1} &
=\hat{1}\\\nonumber
P_{1,-1,1,-1}+P_{1,1,1,1}+P_{1,0,-1,0}+P_{0,1,0,-1} & =\hat{1}\\
P_{0,0,1,0}+P_{0,1,0,0}+P_{1,0,0,1}+P_{1,0,0,-1} &
=\hat{1}\\\nonumber
P_{1,-1,-1,1}+P_{1,1,1,1}+P_{1,0,0,-1}+P_{0,1,-1,0} &
=\hat{1}\\\nonumber
P_{1,1,-1,1}+P_{1,1,1,-1}+P_{1,-1,0,0}+P_{0,0,1,1} &
=\hat{1}\\\nonumber
P_{1,1,-1,1}+P_{-1,1,1,1}+P_{1,0,1,0}+P_{0,1,0,-1} &
=\hat{1}\\\nonumber
P_{1,1,1,-1}+P_{-1,1,1,1}+P_{1,0,0,1}+P_{0,1,-1,0} & =\hat{1},
\end{align}

Each line of Equation \eqref{e:CabelloKS} below represents a
different measurement context. The~subindexes represent
non-normalized vectors in the four-dimensional Hilbert space (each
vector defines a ray, and represents a measurement outcome in that
context). Each measurement context has four orthogonal projections,
that represent the outcomes of a projective measurement (and,
as~such, add up to the identity operator). The~contexts are chosen
in such a way that each pair of them shares one element in common.

This represents the intertwining of the Boolean algebras associated
to the contexts since, for~example, the~event represented by
$P_{0,0,0,1}$ belongs to the first and second contexts. Similarly,
the~event $P_{0,0,1,0}$ belongs to the first and the fifth contexts.
The~family of contexts of this example is chosen in such a way that
each event belongs to exactly two different contexts. Thus, since
there are nine equations, there are eighteen different events in
total.

The~Boolean algebras associated to the nine contexts are related in
a non-trivial way (since the intersection of any two of them is
strictly greater than $\{\mathbf{0},\mathbf{1}\}$). In order to
illustrate the Kochen--Specker theorem, let try to assign truth
values to each of these events, which can be represented as $0$ vs.
$1$ assignments to the different outcomes of the experiments. Thus,
for~example, we can assign $1$ (true) to $P_{0,0,0,1}$, or~$0$
false, and~proceed similarly to the other events.

We represent this by a function $\nu$: $\nu(P_{0,0,0,1})=1$,
$\nu(P_{0,0,1,0})=0$, etc. A~truth value assignment would mean that
each possible experiment outcome has a definite value previous to
measurement. This is related to asking about the existence of a
\emph{dispersion free state}, that is, a~state that only assigns the
probabilities zero and one to all possible outcomes. Thus, the
function $\nu$ must satisfy one condition: given that all the
outcomes in a context are mutually exclusive, the~function must be
defined in such a way that there are no two truth value assignments
in the same context.

Thus, for~example, if~we assign the truth value $1$ to $P_{0,0,0,1}$
(i.e., $\nu(P_{0,0,0,1})=1$), then, all other members of that
context must have  the truth value $0$ assigned
($\nu(P_{0,0,1,0})=\nu(P_{1,1,0,0})=\nu(P_{1,-1,0,0})=0$).
Equation~(\ref{e:CabelloKS}) implies that the valuations must
satisfy $\sum_i \nu(P_{i})=1$ on each line (this is known as the
FUNC condition in the literature; see the discussion and references
in~\cite{Holik-Jorge}). We must assume that the valuations preserve
their values from context to context (if we assign a certain truth
value to a projection in a given context, we must use that same
value when it appears in a different context).

However, it is easy to check that such a compatible truth value
assignment is not possible. The~reason is as follows. There are nine
equations and~eighteen events. If~we sum all equations (of the form
$\sum_i \nu(P_{i})=1$), on~the right, we obtain an odd number
(nine), and~on the left, we obtain an even number, since there is an
even number of ones. However, this is impossible. The~non-existence
of such a truth value assignment shows one of the most important
implications of the intertwining between the Boolean algebras
associated to the contexts.

This is known as the Kochen--Specker theorem
(see~\cite{Cabello-KS-1996} for details). This example illustrates
clearly how the Boolean algebras of events associated to quantum
systems are intertwined and~how this complex structure gives place
to interpretational issues. As~is well known, the Kochen--Specker
theorem is a cornerstone in the discussions about the foundations of
quantum mechanics (see, for example~\cite{Holik-Jorge} and the
references therein).

\end{enumerate}

\subsection{Quantal~Effects}\label{s:QuantalEffects}

Projective measures are not the only way in which observable
quantities can be described in QM. There exists a more general
notion, namely, that of the \emph{quantal effect}. This notion can
be generalized to arbitrary statistical theories. The generalization
of the notion of PVM (which is based on projections) to an
observable based on effects is called a \emph{positive operator
valued measure} (POVM))
\cite{Cattaneo-Gudder-1999,bush,foulis,EffectAlgebras-Foulis-2001,Busch-Lahti-2009,Thesis-Heinonen-2005,Ma-Effects}
and, in QM, will be represented by a mapping
($\mathcal{B}(\mathcal{H})$ stands for the set of bounded operators
in $\mathcal{H}$).
\begin{subequations}\label{e:POVMs}
\begin{equation}\label{e:21a}
E:B(\mathbb{R})\rightarrow\mathcal{B}(\mathcal{H}).
\end{equation}
\noindent such that
\begin{equation}
E(\mathbb{R})=\mathbf{1}
\end{equation}
\begin{equation}
E(B)\geq 0, \,\,\mbox{for any}\,\, B\in B(\mathbb{R})
\end{equation}
\begin{equation}\label{e:21d}
E(\cup_{j}(B_{j}))=\sum_{j}E(B_{j}),\,\, \mbox{for any disjoint
familly}\,\, {B_{j}}
\end{equation}

\end{subequations}

The reader should compare Equations \eqref{e:21a}--\eqref{e:21d}
with \eqref{e:5a}--\eqref{e:5e} and \eqref{e:16a}--\eqref{e:16e}.
A~POVM is, thus, a measure whose values are non-negative
self-adjoint operators on a Hilbert space, and~the above definition
reduces to the PVM case when these operators are also orthogonal
projections. It is the most general formulation of the description
of a measurement in the framework of quantum physics. Positive
operators $E$ satisfying $0\leq E\leq\mathbf{1}$ are called
\emph{effects} and generate an \emph{effect algebra}
\cite{EffectAlgebras-Foulis-2001,Cattaneo-Gudder-1999}). We denote
this algebra by $\mathrm{E}(\mathcal{H})$. It is also important to
remark that POVMs can be associated to fuzzy measurements (and thus
with~fuzzy sets; see~\cite{Ali-POVMs,Thesis-Heinonen-2005}).

In QM, a POVM defines a family of affine functionals on the quantum
state space $\mathcal{C}$ of all positive hermitian trace-class
operators of trace one. Thus, given a Borel set $B$, we have:

\begin{subequations}
\begin{equation}
E(B):\mathcal{C}\rightarrow [0,1]
\end{equation}
\begin{equation}
\rho\mapsto \mbox{tr}(E(B)\rho)
\end{equation}

\end{subequations}

\noindent for every Borel set $B$. This will be relevant in certain
generalizations of quantum probabilities, which we will discuss
below.

\section{Generalization to Orthomodular~Lattices}\label{s:Orthomodular}

In the algebraic formulation of relativistic quantum theory, there
appear algebras that are different from the {ones} used in
non-relativistic QM~\cite{HalvorsonARQFT}. {In the non-relativistic
case, the~algebra $\mathcal{B}(\mathcal{H})$ of all bounded
operators acting on a separable Hilbert space generates---via the
spectral theorem---all possible observables. However, the study of
quantum systems with infinitely many degrees of freedom revealed
that other algebras are needed. Murray and von Neumann provided a
classification of these algebras, which are called Type I, Type II,
and Type III. For~the non-relativistic case with finitely many
degrees of freedom, it suffices to use Type I factors. However,
in~the general case, Type II and Type III factors appear}.

{The existence of different algebraic models of quantum theories
suggests that}, in~principle, one could conceive more general
probabilistic models than those of standard QM. We describe here a
possible generalization, based in orthomodular lattices. Let
$\mathcal{L}$ be an arbitrary orthomodular lattice (standing for the
lattice of all possible empirical events of a given model). Then, we
define

\begin{subequations}\label{e:GeneralizedProbabilityA}
\begin{equation}\label{e:23a}
s:\mathcal{L}\rightarrow [0;1],
\end{equation}

\noindent such that:
\begin{equation}\label{e:Qprobability1}
s(\textbf{1})=1.
\end{equation}

\noindent and, for~a denumerable and pairwise orthogonal family of
events $E_{j}$,
\begin{equation}\label{e:23c}
s(\sum_{j}E_{j})=\sum_{j}s(E_{j}).
\end{equation}
\end{subequations}

If we put $\mathcal{L}=\Sigma$ and
$\mathcal{L}=\mathcal{P}(\mathcal{H})$ in
Equations~\eqref{e:23a}--\eqref{e:23c}, we recover the Kolmogorovian
and quantum cases, respectively. For~a discussion on the conditions
under which measures  those defined in
Equations~\eqref{e:23a}--\eqref{e:23c} are well defined,
see~\cite{belcas81}, Chapter $11$. The~fact that projection
operators of arbitrary von Neumann algebras define orthomodular
lattices~\cite{Redei-Summers2006}, shows that the above
generalization includes many examples of interest (in addition to
classical statistical mechanics and standard QM).

Notice again that the set of all possible measures satisfying
\eqref{e:23a}--\eqref{e:23c} is convex. This opens the door to a
further generalization of probabilistic models based on \emph{convex
sets}, that we discuss in the next~Section.

The states defined in Equations~\eqref{e:23a}--\eqref{e:23c} define
Kolmogorovian probabilities when restricted to maximal Boolean
subalgebras of $\mathcal{L}$. Denote by $\textbf{B}$ to the set of
all possible Boolean subalgebras of $\mathcal{L}$. It is possible to
consider $\textbf{B}$ as a pasting of its maximal Boolean
subalgebras (see, for example~\cite{navara1991pasting} and the
discussions posed in~\cite{HolikQIC-2016,Holik-Entropy-2015}):
\begin{equation}\label{e:BooleanAlgebrasPasting}
\mathcal{L}=\bigvee_{\mathcal{B}\in\textbf{B}}\mathcal{B} \,.
\end{equation}

The decomposition represented by
Equation~\eqref{e:BooleanAlgebrasPasting} implies that a state
defined as a measure over an orthomodular lattice can be considered
as a pasting of Kolmogorovian probabilities. If~there is only one
maximal Boolean subalgebra, then the whole $\mathcal{L}$ has to be
Boolean, and~thus we recover a Kolmogorovian model. In~theories that
display contextuality, such as standard
QM~\cite{Holik-Entropy-2015,HolikQIC-2016,Holik-MachZender}, there
will be more than one empirical context, and~thus the~above
decomposition will not be~trivial.

The representation of observables in this setting can be made as
follows (we follow~\cite{piron} here).

\begin{definition}\label{d:c-morphisms}
A \emph{c-morphism} is a one to one map
$\alpha:\mathcal{L}_{1}\longrightarrow\mathcal{L}_{2}$ between
orthocomplemented complete lattices $\mathcal{L}_{1}$ and
$\mathcal{L}_{2}$ such~that \vspace{-6pt}

\begin{subequations}
\begin{equation}
\alpha(\bigvee_{i}a_{i})=\bigvee_{i}\alpha(a_{i})
\end{equation}
\begin{equation}
a\bot b\Longrightarrow\alpha(a)\bot\alpha(b)
\end{equation}
\begin{equation}
\alpha(\mathbf{1}_{1})=\alpha(\mathbf{1}_{2})
\end{equation}
\end{subequations}
\end{definition}

Given a physical system whose event lattice is given by
$\mathcal{L}$, an~\emph{observable} can be defined as a c-morphism
from a Boolean lattice $\mathcal{B}$ into $\mathcal{L}$:

\begin{definition}[Observable]\label{d:Observable}
An \emph{observable} of a physical system whose event lattice is
$\mathcal{L}$ and that takes its values in the outcome set $M$ will
be a c-morphism $\phi$ from a Boolean algebra $\mathcal{B}_{M}$ of
subsets of $M$, to~a Boolean subalgebra
$\Sigma_{\phi}\subseteq\mathcal{L}$.
\end{definition}

The following diagram illustrates what happens when we compose a
generalized state (given by Equations~\eqref{e:23a}--\eqref{e:23c})
and an observable (as in Definition \eqref{d:Observable}):

\[
\xymatrix{
\mathcal{B}_{M}\ar[d]_{\phi}\ar[r]^-{s}&[0,1]\\
\mathcal{L}\ar@{-->}[ur]^{s\circ\phi}& }
\]

The composition $s\circ\phi:\mathcal{B}_{M}\longrightarrow [0,1]$
defines a Kolmogorovian probability (i.e., a~measure satisfying
Equations~\eqref{e:1a}--\eqref{e:1c}).

Let us now compare compare Equations \eqref{e:5a}--\eqref{e:5e} and
\eqref{e:16a}--\eqref{e:16e} with Definitions \eqref{d:c-morphisms}
and \eqref{d:Observable}. By~looking at the definition of PVM
(Equations \eqref{e:16a}--\eqref{e:16e}), it is easy to recognize
that a PVM is a c-morphism between the set of Borel subsets of
$\mathcal{B}(\mathbb{R})$ and the Boolean algebra generated by its
image projections. According to the above definition of observable,
one can quickly realize that any Boolean subalgebra of $\mathcal{L}$
will determine an observable (more properly, a~family of observables
up to rescaling). For~the classical case, by~looking again at the
``important remark'' of Section \ref{s:RandomVariables} (Equations
\eqref{e:5a}--\eqref{e:5e}), we realize that a classical random
variable also satisfies the general definition of observable given
in Definition \ref{d:Observable}.

\section{Convex Operational~Models}\label{s:COMpreliminaries}

In the previous section, we demonstrated that the set of states
defined over an arbitrary orthomodular lattice is convex. This
approach contains the quantum and classical state spaces as
particular cases. Thus, it seems very natural to attempt to define
generalized probabilistic models by appealing to convex sets.

This key observation leads to a general approach to statistical
theories based on the study of the geometrical properties of convex
sets. This is the starting point of the \textit{Convex Operational
Models} (COM) approach. In~this section, we concentrate on
elementary notions of COM's, and~we refer the reader
to~\cite{Barnum-PRL} for an excellent presentation of the subject.
The~approach based on convex sets results as more general than the
one based in orthomodular lattices (i.e., the~latter can be included
as particular cases of the COM approach).

If the state space of a given probabilistic theory is given by the
set $\mathbf{S}$, let us denote by $X$ to the set of possible
measurement outcomes of an observable quantity. Then, if~the system
is in a state $s$, a~probability $p(x,s)$ is assigned to any
possible outcome $x\in X$. This probability should be well defined
in order that our theory be considered as a probabilistic one.
In~this way, we must have a function
\begin{eqnarray}
&p:X\times\mathbf{S}\mapsto [0,1]&\nonumber\\
&(x,s)\rightarrow p(x,s)&
\end{eqnarray}

To each outcome $x\in X$ and state $s\in\mathbf{S}$, this function
assigns a probability $p(x,s)$ of $x$ to occur if the system is in
the state $s$. In~this way, a~triplet
$(\mathbf{S},p(\cdotp,\cdotp),X)$ is assigned for each system of any
probabilistic theory~\cite{mackey-book}. Thinking of $s$ as a
variable, we obtain a mapping $s\mapsto p(\cdot,s)$ from
$\mathbf{S}\rightarrow [0,1]^{X}$. This implies that all the states
of $\mathbf{S}$ can be identified with maps, which generates a
canonical vector space. Their closed convex hull forms a new set
$\mathcal{S}$ representing all possible probabilistic mixtures
(convex combinations) of states in $\mathbf{S}$. Given an arbitrary
$\alpha\in\mathcal{S}$ and any outcome $x\in X$, we can define an
affine evaluation-functional $f_{x}:\mathcal{S}\rightarrow [0,1]$ in
a canonical way by $f_{x}(\alpha)=\alpha(x)$.

More generally, we can consider any affine functional
$f:\mathcal{S}\rightarrow [0,1]$ as representing a measurement
outcome, and~thus use $f(\alpha)$ to represent the probability for
that outcome in state $\alpha$. We will call $A(\mathcal{S})$ to the
space of all affine functionals. Due to the fact that QM is also a
probabilistic theory, it follows that it can be included in the
general framework described above (we denoted its convex set of
states by $\mathcal{C}$ in Section~\ref{s:QuantumProbabilities}).
In~QM, affine functionals defined as above are called
\textit{effects} (and are coincident with the constituents of POVM's
as defined in Section~\ref{s:QuantalEffects}). The~generalized
probabilistic models defined in Section~\ref{s:Orthomodular} fall
naturally into the scope of the COM approach, given that their state
spaces are convex sets.

We saw that a probability $a(\omega)\in[0,1]$ is well defined for
any state $\omega\in\mathcal{S}$ and an observable $a$. In~the COM
approach, it is usually assumed that there exists a unitary
observable $u$ such that $u(\omega)=1$ for all
$\omega\in\mathcal{S}$. Thus, in~analogy with the quantum case,
the~set of all effects will be encountered in the interval $[0,u]$
(the order in the general case is the canonical one in the space of
affine functionals). A~(discrete) measurement will be represented by
a set of effects $\{a_{i}\}$ such that $\sum_{i}a_{i}= u$.
$\mathcal{S}$ can be naturally embedded in the dual space
$A(\mathcal{S})^{\ast}$ using the map
\begin{eqnarray}
\omega\mapsto\hat{\omega}\nonumber\\
\quad \hat{\omega}(a):=a(\omega)
\end{eqnarray}

Let $V(\mathcal{S})$ be the linear span of $\mathcal{S}$ in
$A(\mathcal{S})^{\ast}$. Then, it is reasonable to consider
$\mathcal{S}$ finite dimensional if and only if $V(\Omega)$ is
finite dimensional. For~the sake of simplicity, we restrict
ourselves to this case (and to compact spaces). As~is well known,
this implies that $\mathcal{S}$ can be expressed as the convex hull
of its extreme points. The~extreme points will represent \emph{pure
states} (in the QM case, pure quantum states are indeed the extreme
points of $\mathcal{C}$, and~correspond to one dimensional
projections in the Hilbert space).

It can be shown that, for finite dimensions $d$, a~system will be
classical if and only if it is a simplex (a simplex is the convex
hull of $d+1$ linearly independent pure states). It is a well known
fact that, in a simplex, a~point may be expressed as a unique convex
combination of its extreme points. This characteristic feature of
classical theories no longer holds in quantum models. Indeed, in~the
case of QM, there are infinite ways in which one can express a mixed
state as a convex combination of pure states (for a graphical
representation, think about the maximally mixed state in the Bloch
sphere).

Interestingly enough, there is also a connection between the faces
of the convex set of states of a given model and its lattice of
properties (in the quantum-logical sense), providing an unexpected
connection between geometry, lattice theory, and statistical
theories. Faces of a convex set are defined as subsets that are
stable under mixing and purification. This is means that a convex
subset F is a face if, each time that
\begin{equation}
x=\lambda x_1+(1-\lambda)x_2, \,\,\,\,\,0\leq\lambda\leq1,
\end{equation}

\noindent and then $x\in F$ if and only if $x_1\in F$ and $x_2\in F$
\cite{BEN}. The~set of faces of a convex set forms a lattice in a
canonical way, and it can be shown that the lattice of faces of a
classical model is a Boolean one. On~the other hand, in~QM,
the~lattice of faces of the convex set of states $\mathcal{C}$
(defined as the set of positive trace class hermitian operators of
trace one) is isomorphic to the von Neumann lattice of closed
subspaces $\mathcal{P}(\mathcal{H})$ \cite{BEN,belcas81}. For~a
general model, the lattice of faces may fail to be suitably
orthocomplemented~\cite{belcas81} (and thus the~COM approach is more
general than the one based in orthomodular lattices).

Let us turn now to compound systems. Given a compound system, its
components will have state spaces $\mathcal{S}_{A}$ and
$\mathcal{S}_{B}$. Let us denote the joint state space by
$\mathcal{S}_{AB}$ . It is reasonable to identify $\mathcal{S}_{AB}$
with the linear span of $(V(\mathcal{S}_{A})\otimes
V(\mathcal{S}_{B}))$ \cite{Barnum-PRL}. Then, a~maximal tensor
product state space $\mathcal{S}_{A}\otimes_{max}\mathcal{S}_{B}$
can be defined as  one that contains all bilinear functionals
$\varphi:A(\mathcal{S}_{A})\times
A(\mathcal{S}_{B})\longrightarrow\mathbb{R}$ such that
$\varphi(a,b)\geq 0$ for all effects $a$ and $b$ and
$\varphi(u_{A},u_{B})=1$. The~maximal tensor product state space has
the property of being the largest set of states in
$(A(\mathcal{S}_{A})\otimes A(\mathcal{S}_{B}))^{\ast}$, which
assigns probabilities to all product-measurements. The~minimal
tensor product state space
$\mathcal{S}_{A}\otimes_{min}\mathcal{S}_{B}$ is simply defined by
the convex hull of all product states. A~product state will then be
a state of the form $\omega_{A}\otimes\omega_{B}$ such that
\begin{equation}
\omega_{A}\otimes\omega_{B}(a,b)=\omega_{A}(a)\omega_{B}(b),
\end{equation}

\noindent for all pairs $(a,b)\in A(\mathcal{S}_{A})\times
A(\mathcal{S}_{B})$. Given a particular compound system of a general
statistical theory, its set of states $\mathcal{S}_{AB}$---we call
it $\mathcal{S}_{A}\otimes\mathcal{S}_{B}$ from now on---will
satisfy
\begin{equation}\label{e:Inclusions}
\mathcal{S}_{A}\otimes_{min}\mathcal{S}_{B}\subseteq\mathcal{S}_{A}\otimes\mathcal{S}_{B}\subseteq\mathcal{S}_{A}\otimes_{max}\mathcal{S}_{B}
\end{equation}

\noindent As expected, for~classical compound systems (because of
the absence of entangled states), we have
$\mathcal{S}_{A}\otimes_{min}\mathcal{S}_{B}=\mathcal{S}_{A}\otimes_{max}\mathcal{S}_{B}$.
In the quantum case, we have strict inclusions
in~\eqref{e:Inclusions}:
$\mathcal{S}_{A}\otimes_{min}\mathcal{S}_{B}\subsetneq\mathcal{S}_{A}\otimes\mathcal{S}_{B}\subsetneq\mathcal{S}_{A}\otimes_{max}\mathcal{S}_{B}$.
The general definition of a separable state in an arbitrary COM is
made in analogy with that of~\cite{Werner}, i.e.,~as one that can be
written as a convex combination of product
states~\cite{Barnum-Toner,Perinotti-2011} (see
also~\cite{Holik-Plastino-Massri} for a generalization):

\begin{definition}\label{d:generalseparable}
A state $\omega\in\mathcal{S}_{A}\otimes\mathcal{S}_{B}$ will be
called \emph{separable} if there exist $p_{i}\in\mathbb{R}_{\geq
0}$, $\omega^{i}_{A}\in\mathcal{S}_{A}$ and
$\omega^{i}_{B}\in\mathcal{S}_{B}$ such that
\begin{equation}
\omega=\sum_{i}p_{i}\omega^{i}_{A}\otimes\omega^{i}_{B},\quad \sum_i
p_i =1
\end{equation}

\end{definition}

If $\omega\in\mathcal{S}_{A}\otimes\mathcal{S}_{B}$ is not
separable, then it will be reasonably called \emph{entangled}
\cite{BEN,Schro1}.
As~expected, entangled states exist only if
$\mathcal{S}_{A}\otimes\mathcal{S}_{B}$ is strictly greater than
$\mathcal{S}_{A}\otimes_{min}\mathcal{S}_{B}$.

The COM approach already shows that, given an arbitrary statistical
theory, there is a generalized notion of probabilities of
measurement outcomes. These probabilities are encoded on the states
in $\mathcal{S}$. We have seen that there are many differences
between classical state spaces and non-classical ones: this is
expressed in the geometrical properties of their convex state spaces
and in the correlations appearing when compound systems are
considered. Indeed, QM and classical probability theories are just
particular COMs among a vast family of~possibilities.

It is important to remark that many informational techniques, such
as the MaxEnt method, can be suitably generalized to arbitrary
probabilistic
models~\cite{Holik-JaynesGroup,Holik-Plastino-QuantalEffects-2012}.
In~a similar vein, quantum information theory could be considered as
a particular case of a generalized information
theory~\cite{Holik-Entropy-2015}.

\section{Cox's Method Applied To~Physics}\label{s:KnuthGoyal}

Now, we review a relatively recent approach to the probabilities
appearing in QM that uses distributive lattices. A~novel derivation
of Feynman's rules for quantum mechanics was presented in
~\cite{GoyalKnuthSkilling,Symmetry}. There, an~experimental
\emph{logic of processes} for quantum systems is presented, and~this
is done in such a way that the resulting lattice is a distributive
one. This is a major difference with the approach described in
Section~\ref{s:QuantumProbabilities} because~the lattice of
projections in a Hilbert space is~non-distributive.

The logic of processes is constructed as follows. Given a sequence
of measurements $M_1$,$\ldots$,$M_n$ on a quantum system, yielding
results $m_1$, $m_2$, $\ldots$, $m_n$, a~particular process is
represented as a \emph{measuring sequence} $A=[m_1,m_2,\ldots,m_n]$.

Next, conditional (logical) propositions $[m_2,\ldots,m_n|m_1]$ are
introduced. Using them, a~probability is naturally associated to a
sequence $A$ with the formula
\begin{equation}
P(A)=Pr(m_n,\ldots,m_2|m_1)
\end{equation}

\noindent representing the probability of obtaining outcomes $m_2$,
$\ldots$, $m_n$ conditional upon obtaining $m_1$.

Let us see how this works with a concrete example in which the
$m_i$'s have two possible values, $0$ and $1$. Then, $A_1=[0,1,1]$
and $A_2=[0,0,1,1]$ represent the measurement sequences of three and
four measurements, respectively. Here, $P(A_1)=Pr(1,1|0)$ represents
the probability of obtaining outcomes $m_2=1$ and $m_3=1$
conditional upon obtaining $m_1=0$.

Measurements can be coarse grained as follows. Suppose that we want
to coarse grain $M_2$. Then, we can unite the two outcomes $0$ and
$1$ in a joint outcome $(0,1)$. Then, a~new measurement
$\widetilde{M}_2$ is created. Thus, a~possible sequence obtained by
the replacement of $M_2$ by $\widetilde{M}_2$ could be
{$[1,(0,1),1]$}. Analogous constructions can be done for other
measurements. In~this way, an operation can be defined for the
sequences:
\begin{equation}
[m_1,\ldots,(m_i,m'_i),\ldots,m_n]:=[m_1,\ldots,m_i,\ldots,m_n]\vee[m_1,\ldots,m'_i,\ldots,m_n]
\end{equation}

Another operation can be defined reflecting the fact that sequences
can be compounded as follows
\begin{equation}
[m_1,\ldots,m_j,\ldots,m_n]:=[m_1,\ldots,m_j]\cdot[m_j,\ldots,m_n]
\end{equation}

\noindent Notice that, in~the above equation, the~last measurement
and outcome of the first sequence must be the same as the first
measurement and outcome of the second. With~these operations at
hand, it is easy to show that, if $A$, $B$, and $C$ are measuring
sequences, then

\begin{subequations}\label{e:ExperimentalLogicKnuth}
\begin{equation}\label{e:35a}
A\vee B=B\vee A
\end{equation}
\begin{equation}
(A\vee B)\vee C=A\vee(B\vee C)
\end{equation}
\begin{equation}
(A\cdot B)\cdot C=A\cdot(B\cdot C)
\end{equation}
\begin{equation}
(A\vee B)\cdot C=(A\cdot C)\vee(B\cdot C)
\end{equation}
\begin{equation}\label{e:35e}
C\cdot(A\vee B)=(C\cdot A)\vee(C\cdot B),
\end{equation}

\end{subequations}

Equations \eqref{e:35a}--\eqref{e:35e} show explicitly that
``$\vee$'' is commutative and associative, ``$\cdot$'' is
associative, and that there is right- and left-distributivity of
``$\cdot$'' over ``$\vee$''.

Equations \eqref{e:35a}--\eqref{e:35e} define the algebraic
``symmetries'' of the experimental logic of processes. As~in the
approach of Cox to classical probability, these symmetries are used
to derive Feynman's rules~\cite{Symmetry}. However, at this step,
a~crucial assumption is made: each measuring sequence will be
represented by a pair of real numbers~\cite{Symmetry}. This
assumption is justified in~\cite{Symmetry} by appealing to Bohr's
complementarity~principle.

If measuring sequences $A$, $B$, etc. induces pairs of real numbers
$\mathbf{a}$, $\mathbf{b}$, etc., then, due to
Equations~\eqref{e:35a}--\eqref{e:35e}, {the associated pairs should
satisfy}

\begin{subequations}\label{e:ExperimentalLogicComplex}
\begin{equation}\label{e:36a}
\mathbf{a}\vee \mathbf{b}=\mathbf{b}\vee \mathbf{a}
\end{equation}
\begin{equation}
(\mathbf{a}\vee \mathbf{b})\vee
\mathbf{c}=\mathbf{a}\vee(\mathbf{b}\vee \mathbf{c})
\end{equation}
\begin{equation}
(\mathbf{a}\cdot
\mathbf{b})\cdot\mathbf{c}=\mathbf{a}\cdot(\mathbf{b}\cdot
\mathbf{c})
\end{equation}
\begin{equation}
(\mathbf{a}\vee \mathbf{b})\cdot \mathbf{c}=(\mathbf{a}\cdot
\mathbf{c})\vee(\mathbf{b}\cdot \mathbf{c})
\end{equation}
\begin{equation}\label{e:36e}
\mathbf{c}\cdot(\mathbf{a}\vee \mathbf{b})=(\mathbf{c}\cdot
\mathbf{a})\vee(\mathbf{c}\cdot \mathbf{b})
\end{equation}

\end{subequations}

The reader can easily verify that Equations
\eqref{e:36a}--\eqref{e:36e} are satisfied by the field of complex
numbers (provided that the operations are interpreted as sum and
product of complex numbers). How can we be assured that complex
numbers are the only field that satisfies Equations
\eqref{e:36a}--\eqref{e:36e}? In order to single out complex numbers
among other possible fields, additional assumptions must be added,
namely, pair symmetry, additivity, and symmetric bias
(see~\cite{Symmetry,GoyalKnuthSkilling} for details). Once these
conditions are assumed, the~path is clear to derive Feynman's rules
by applying a deduction similar to that of Cox, to~the experimental
logic defined by Equations \eqref{e:35a}--\eqref{e:35e}.

\section{Generalization of Cox's Method}\label{s:QMDerivation}

As we have seen in previous sections, there are two versions of CP,
namely, the~approach of R. T. Cox~\cite{CoxPaper,CoxLibro} and the
one of A. N. Kolmogorov~\cite{KolmogorovProbability}.
The~Kolmogorovian approach can be generalized in order to include
non-Boolean models, as~we have shown in
Section~\ref{s:Orthomodular}. In~what follows, we will see that
Cox's method can also be generalized to non-distributive lattices,
and~thus the~non-commutative character of QP can be captured in this
framework~\cite{HolikRingsOfOperators,Holik-Plastino-Saenz-2012}.

\subsection*{Generalized Probability Calculus Using Cox's Method}

As we have seen in Section~\ref{s:CoxReview}, Cox studied the
functions defined over a distributive lattices and derived classical
probabilities. In~\cite{Holik-Plastino-Saenz-2012}, it is shown that
if the lattice is assumed to be non-distributive, the~properties of
QP described in Section~\ref{s:QuantumProbabilities} can be derived
by applying a variant of Cox's method as follows
(see~\cite{Holik-Plastino-Saenz-2012} for details). Suppose that the
propositions of our system are represented by the lattice of
elementary tests of QM, i.e.,~the lattice of projections
$\mathcal{P}(\mathcal{H})$ of the Hilbert space $\mathcal{H}$.
The~goal is to show that the ``degree of implication'' measure
$s(\cdots)$ demanded by Cox's method satisfies
Equations~\eqref{e:17a}--\eqref{e:17c}. This means that we are
looking for a function to the real numbers $s$, such that it is
non-negative and $s(P)\leq s(Q)$ whenever $P\leq Q$.

The operation ``$\vee$'' in $\mathcal{P}(\mathcal{H})$ is
associative. Then, if~$P$ and $Q$ are orthogonal projections,
the~relationship between $s(P)$, $s(Q),$ and $s(P\vee Q)$ must be of
the form
\begin{equation}\label{e:s(or)}
s(P\vee Q)=F(s(P),s(Q)),
\end{equation}

\noindent with $F$ a function to be determined. {If a third
proposition $R$ is added, following a similar procedure to that of
Cox, we obtain for ``$P\vee P\vee R$'' the following functional
equation}
\begin{equation}\label{e:FunctEq1}
F(F(s(P),s(Q)),s(R))=F(s(P),F(s(Q),s(R))).
\end{equation}

The above equation can be solved up to
rescaling~\cite{aczel-book,Knuth-2004a,Knuth-2004b,Knuth-2005b},
and~we find
\begin{equation}
s(P\vee Q)=s(P)+s(Q).
\end{equation}

\noindent whenever $P\perp Q$. It can be shown that, for any finite
family of orthogonal projections $P_j$, $1\leq j\leq n$
\cite{Holik-Plastino-Saenz-2012}:
\begin{equation}
s(\bigvee_{j=1}^{\infty}P_j)=\sum_{j=1}^{\infty}s(P_j),
\end{equation}

\noindent and we  recover condition \eqref{e:23c} of the axioms of
quantum probability. By~exploiting the properties of the orthogonal
complement acting on subspaces, it can also be
shown~\cite{Holik-Plastino-Saenz-2012}~that
\begin{equation}
s(P^{\perp})=1-s(P),
\end{equation}

On the other hand, as~$\mathbf{0}=\mathbf{0}\vee\mathbf{0}$ and
$\mathbf{0}\bot\mathbf{0}$, then
$s(\mathbf{0})=s(\mathbf{0})+s(\mathbf{0})$, and~thus,
$s(\mathbf{0})=0$, which is condition \eqref{e:Qprobability1}.
In~this way, it follows that Cox's method applied to the
non-distributive lattice $\mathcal{P}(\mathcal{H})$ yields the same
probability theory as the one provided by Equations
\eqref{e:17a}--\eqref{e:17c} for the quantum~case.

What happens if Cox's method is applied to an arbitrary atomic
orthomodular complete lattice $\mathcal{L}$? Now, we must define a
function $s:\mathcal{L}\longrightarrow \mathbb{R}$, such that it is
always non-negative $s(a)\geq 0\,\,\forall a\in\mathcal{L}$ and is
also order preserving $a\leq b \longrightarrow s(a)\leq s(b)$.
In~\cite{Holik-Plastino-Saenz-2012}, it is shown that, under these
rather general assumptions, in~\emph{any} atomic orthomodular
lattice and for any orthogonal denumerable family
$\{a_i\}_{i\in\mathbb{N}}$, $s$ must satisfy (up to rescaling)

\begin{subequations}\label{e:GeneralizedProbability}
\begin{equation}\label{e:42a}
s(\bigvee\{a_i\}_{i\in\mathbb{N}})=\sum_{i=1}^{\infty}s(a_i)
\end{equation}
\begin{equation}
s(\neg a)= 1-s(a)
\end{equation}
\begin{equation}\label{e:42c}
s(\mathbf{0})=0.
\end{equation}
\end{subequations}

In this way, a~generalized probability theory is derived (as in
\eqref{e:17a}--\eqref{e:17c}).
Equations~\eqref{e:42a}--\eqref{e:42c} define non-classical
(non-Kolmogorovian) probability measures, due to the fact that, in
any non-distributive orthomodular lattice, there always exist
elements $a$ and $b$ such that
\begin{equation}
(a\wedge b)\vee (a\wedge\neg b)<a,
\end{equation}

\noindent However, in~any classical probability theory,
$s(a\wedge\neg b)+s(a\wedge b)= s(a)$ is always~satisfied.

In~the non-Boolean setting of QM, von Neumann's entropy (VNE) plays
a similar role to that of Shannon's in Cox
approach~\cite{HolikQIC-2016}. This allows us to interpret the VNE
as a natural measures of information for an experimenter who deals
with a contextual event~structure.

\section{Conclusions}
\label{s:Conclusions}

We  presented a new approach for probabilities appearing in QM.
While there exist (at least) two alternative formalisms to CP (the
Kolmogorovian and the one due to R. T. Cox), we have also shown that
these two approaches can be extended to the non-commutative case.
In~this way, we find that CP are a particular case of a more general
mathematical framework in which the lattice is distributive. QP is
also a particular case of a vast family of theories for which the
propositional lattice is non-distributive. Thus, we have a precise
formal expression of the notion of~QP.

These formal frameworks do not exhaust the philosophical debate
around the existence or not of a well-defined notion of QP;
notwithstanding, the~extension of Cox's method to the
non-distributive case, as well as the possibility of including a
description of the probabilities in QM in it, constitutes a precise
step towards understanding the notion of QP, offering a new point of
view of this notion. According to this interpretation, a~rational
agent is confronted with a particular event structure. To~fix ideas,
suppose that the agent is confronted with a physical system,
and~that the agent has to perform experiments and determine degrees
of belief about their possible outcomes.

\begin{itemize}
\item If the lattice of events that the agent is facing is Boolean (as
in Cox's approach), then, the~measures of degree of belief will obey
laws equivalent to those of Kolmogorov.
\item On the contrary, if~the state of affairs that the agent must face
presents contextuality (as in standard quantum mechanics),
the~measures involved must be
non-Kolmogorovian~\cite{Holik-Plastino-Saenz-2012}.
\item Random variables and information measures~\cite{HolikQIC-2016}
will be the natural generalizations of the classical case if the
event structure is not classical. A~similar observation holds for
the application of the MaxEnt
method~\cite{Holik-JaynesGroup,Holik-Plastino-QuantalEffects-2012}.
\end{itemize}

Our approach allows for a natural justification of the peculiarities
arising in quantum phenomena from the standpoint of a Bayesian
approach. In~particular, quantum information theory could be
considered as a non-Kolmogorovian extension of Shannon's
theory~\cite{Holik-Entropy-2015}. Our approach can be considered as
an alternative step to address Hilbert's problem for the case of
probability theory in QM: the development of an axiomatization
endowed with a clear and natural interpretation of the notions
involved.


\vspace{6pt}
%

This work was partially supported by the grants PIP N$^\circ$
6461/05 amd 1177 (CONICET). In addition, by the projects
FIS2008-00781/FIS (MICINN)---FEDER (EU) (Spain, EU). F.H. was
partially funded by the project ``Per un'estensione semantica della
Logica Computazionale Quantistica-Impatto teorico e ricadute
implementative'', Regione Autonoma della Sardegna, (RAS:
RASSR40341), L.R. 7/2017, annualit\`{a} 2017---Fondo di Sviluppo e
Coesione (FSC) 2014-2020 and the Project PICT-2019-01272.


\appendix
\section{Lattice Theory}\label{s:LatticeTheory}

Lattices can be defined by using equations, i.e.,~they can be
characterized as algebraic structures satisfying certain axiomatic
identities. A~set $\mathcal{L}$ endowed with two operations $\wedge$
and $\vee$ will be called a \emph{lattice}, if,~for all
$x,y,z\in\mathcal{L}$, the following equations are~satisfied.

\begin{subequations}\label{e:LatticeEquations}
\begin{equation}
x\vee x = x\,,\,\, x\wedge x = x\,\,\, \mbox{(idempotence)}
\end{equation}
\begin{equation}
x\vee y = y\vee x\,,\,\, x\wedge y = y\wedge x \,\,\,
\mbox{(commutativity)}
\end{equation}
\begin{equation}
x\vee (y\vee z) = (x\vee y)\vee z \,,\,\, x\wedge (y\wedge z) =
(x\wedge y)\wedge z \,\,\, \mbox{(associativity)}
\end{equation}
\begin{equation}
x\vee (x\wedge y) = x\wedge (x\vee y) = x \,\,\, \mbox{(absortion)}
\end{equation}

\end{subequations}
\noindent If the extra~relationships

\begin{subequations}\label{e:LatticeEquationsDistributive}
\begin{equation}
x\wedge (y\vee z) = (x\wedge y) \vee (x\wedge z)
\,\,\,\mbox{(distributivity 1)}
\end{equation}
\begin{equation}
x\vee (y\wedge z) = (x\vee y) \wedge (x\vee z)\,\,\,
\mbox{(distributivity 2)}
\end{equation}
\end{subequations}
\noindent are satisfied, the~lattice is called \emph{distributive}.

Lattice theory can also be studied using \textit{partially ordered
sets} (\textit{poset}). A~poset is a set $X$ endowed with a partial
ordering relation ``$<$''~satisfying

\begin{itemize}

\item For all $x,y\in X$, if~$x<y$ and $y<x$, then $x=y$.

\item For all $x,y,z\in X$, if~$x<y$ and $y<z$, then $x<z$.

\end{itemize}

We use the notation ``$x\leq y$'' for the case ``$x<y$'' or
``$x=y$''. A lattice $\mathcal{L}$ will be a poset for which any two
elements $x$ and $y$ have a unique supremum and a unique infimum
with respect to the order structure. The~least upper bound of two
given elements ``$x\vee y$'' is called the ``join'', and their
greatest lower bound ``$x\wedge y$'' is called their ``meet''.

A~lattice for which all its subsets have both a supremum and an
infimum is called a \emph{complete lattice}. If,~furthermore, there
exists a greatest element $\mathbf{1}$ and a least element
$\mathbf{0}$, the~lattice is called \emph{bounded}. They are usually
called the \emph{maximum} and the \emph{minimum}, respectively. Any
lattice can be extended into a bounded lattice by adding a greatest
and a least element. Every non-empty finite lattice is bounded.
Complete lattices are always bounded. An~orthocomplementation in a
bounded poset $P$ is a unary operation ``$\neg$'' satisfying:

\begin{subequations}\label{e:ComplementationAxioms}
\begin{equation}\label{e:Complement1}
\neg(\neg(a))=a
\end{equation}
\begin{equation}\label{e:Complement2}
a\leq b \longrightarrow \neg b\leq \neg a
\end{equation}
$a\vee \neg a$ and $a\wedge \neg a$ exist and
\begin{equation}\label{e:Complement3}
a\vee \neg a=\mathbf{1}
\end{equation}
\begin{equation}\label{e:Complement4}
a\wedge \neg a=\mathbf{0}
\end{equation}
hold.
\end{subequations}

A bounded poset with orthocomplementation will be called an
\emph{orthoposet}. An~\emph{ortholattice}, will be an orthoposet,
which is also a lattice. For~$a,\,b \in \mathcal{L}$ (an
ortholattice or orthoposet), we say that $a$ is orthogonal to $b$ ($
a \bot b$) if $a\le \neg b$.  Following~\cite{RedeiHandbook}, we
define an \textit{orthomodular lattice} as an ortholattice
satisfying the orthomodular law:
\begin{equation}\label{e:ModularIdentity}
a \leq b \Longrightarrow a \vee (\neg a \wedge b) = b
\end{equation}
\noindent A \textit{modular lattice}, is an ortholattice satisfying
the stronger condition (modular law)
\begin{equation}\label{e:ModularIdentity2}
a \leq b \Longrightarrow a \vee (b \wedge c) = (a \vee b) \wedge
(a\vee c),
\end{equation}
\noindent and finally a~\textit{Boolean lattice} will be an
ortholattice satisfying the still stronger condition (distributive
law)
\begin{equation}\label{e:DistributiveLaw}
a \vee (b \wedge c) = (a \vee b) \wedge (a\vee c)
\end{equation}

\noindent Thus, a~\emph{Boolean lattice} is a \textbf{complemented
distributive lattice}. We use the terms \emph{Boolean lattice} and
\emph{Boolean algebra} interchangeably.

If $\mathcal{L}$ has a null element $ 0$, then an element $x$ of
$\mathcal{L}$ is an {\it atom} if $0 < x$ and there exists no
element $y$ of $\mathcal{L}$ such that $0 < y < x$. $\mathcal{L}$ is
said to~be:

\begin{itemize}
\item {\it Atomic}, if,~for every nonzero element $x$ of $\mathcal{L}$,
there exists an atom $a$ of $\mathcal{L}$ such that $ a \leq x$.
\item Atomistic, if~every element of $\mathcal{L}$ is a supremum of
atoms.
\end{itemize}

\begin{center}
\begin{tabular}{ c c c }
 \textbf{Boolean} & \textbf{Modular} & \textbf{Orthomodular} \\
 $x\wedge(y\vee z)=(x\wedge y)\vee(x\wedge z)$&$a\leq b\Longrightarrow a\vee(b\wedge c)=(a\vee b)\wedge
(a\vee c)$ & $a \leq b \Longrightarrow a \vee (\neg a \wedge b) = b$\\
 $x\vee (y\wedge z) = (x\vee y) \wedge (x\vee z)$ &  & \\
\end{tabular}
\end{center}


\end{document}